\documentclass[12pt]{article}

\usepackage{amsmath}
\usepackage{amsfonts}
\usepackage{amssymb}

\usepackage{graphicx}

\renewcommand{\theequation}{\arabic{equation}}
\newcommand{\EQ}{\begin{equation}}
\newcommand{\EN}{\end{equation}}

\newcommand{\bear}{\begin{eqnarray}}
\newcommand{\ear}{\end{eqnarray}}
\newcommand{\bt} { \begin{tabular} }
\newcommand{\et}{ \end{tabular} }
\newcommand{\bc} { \begin{center} }
\newcommand{\ec}{ \end{center} }

\newcommand{\btb} { \begin{table} }
\newcommand{\etb}{ \end{table} }

\newcommand{\dpp}{\Delta_{+}}
\newcommand{\dmm}{\Delta_{-}}
\newcommand{\dpm}{\Delta_{\pm}}

\newcommand{\ga}{\gamma}

\newcommand{\OM}{\Omega}

\newcommand{\GAP}{\Gamma_{+}}
\newcommand{\GAM}{\Gamma_{-}}
\newcommand{\GAPM}{\Gamma_{\pm}}

\newcommand{\LAP}{\Lambda_{+}}
\newcommand{\LAM}{\Lambda_{-}}
\newcommand{\LAPM}{\Lambda_{\pm}}

\newcommand{\la}{\lambda}

\newcommand{\TP}{\Theta_{+}}
\newcommand{\TM}{\Theta_{-}}
\newcommand{\TPM}{\Theta_{\pm}}

\newcommand{\DGP}{\Delta_{+}^{(g)}}
\newcommand{\DGM}{\Delta_{-}^{(g)}}
\newcommand{\DGPM}{\Delta_{\pm}^{(g)}}

\newcommand{\DHP}{\Delta_{+}^{(h)}}
\newcommand{\DHM}{\Delta_{-}^{(h)}}
\newcommand{\DHPM}{\Delta_{\pm}^{(h)}}

\newcommand{\DZHP}{\Delta_{+}^{(\tilde{h})}}
\newcommand{\DZHM}{\Delta_{-}^{(\tilde{h})}}
\newcommand{\DZHPM}{\Delta_{\pm}^{(\tilde{h})}}

\newcommand{\EPS}{\epsilon}

\begin{document}

\topmargin 0pt
\oddsidemargin 5mm
\newcommand{\NP}[1]{Nucl.\ Phys.\ {\bf #1}}
\newcommand{\PL}[1]{Phys.\ Lett.\ {\bf #1}}
\newcommand{\NC}[1]{Nuovo Cimento {\bf #1}}
\newcommand{\CMP}[1]{Comm.\ Math.\ Phys.\ {\bf #1}}
\newcommand{\PR}[1]{Phys.\ Rev.\ {\bf #1}}
\newcommand{\PRL}[1]{Phys.\ Rev.\ Lett.\ {\bf #1}}
\newcommand{\MPL}[1]{Mod.\ Phys.\ Lett.\ {\bf #1}}
\newcommand{\JETP}[1]{Sov.\ Phys.\ JETP {\bf #1}}
\newcommand{\TMP}[1]{Teor.\ Mat.\ Fiz.\ {\bf #1}}

\renewcommand{\thefootnote}{\fnsymbol{footnote}}

\newpage
\setcounter{page}{0}
\begin{titlepage}
\begin{flushright}
UFSCARF-TH-10-10
\end{flushright}
\vspace{0.5cm}
\begin{center}
{\large The Yang-Baxter Equation for $\mathcal{PT}$ Invariant Nineteen Vertex Models}\\
\vspace{1cm}
{\large R.A. Pimenta and M.J. Martins} \\
\vspace{0.5cm}
{\em Universidade Federal de S\~ao Carlos\\
Departamento de F\'{\i}sica \\
C.P. 676, 13565-905~~S\~ao Carlos(SP), Brazil}\\
\end{center}
\vspace{0.5cm}

\begin{abstract}
We study the solutions of the Yang-Baxter 
equation associated to nineteen vertex models invariant
by the parity-time symmetry from the perspective of algebraic geometry. We determine the form of
the algebraic curves constraining the respective Boltzmann weights and found that they  
possess a universal structure.
This allows us to classify the integrable manifolds in four different
families reproducing three known models besides uncovering a novel
nineteen vertex model  
in a unified way. The introduction of the spectral parameter on the weights
is made via the parameterization of the fundamental algebraic curve which is a conic. The
diagonalization of the transfer matrix of the new vertex model and its
thermodynamic limit properties are discussed.
We point out a connection between the form of the main curve and the nature
of the excitations of the corresponding spin-1 chains.
\end{abstract}

\vspace{.15cm}
\centerline{PACS numbers:  05.50+q, 02.30.IK}
\vspace{.1cm}
\centerline{Keywords: Yang-Baxter Equation, Lattice Integrable Models, Bethe Ansatz}
\vspace{.15cm}
\centerline{October 2010}

\end{titlepage}


\pagestyle{empty}

\newpage

\pagestyle{plain}
\pagenumbering{arabic}

\renewcommand{\thefootnote}{\arabic{footnote}}

\section{Introduction}

An important concept in the theory of soluble 
two-dimensional lattice models of statistical mechanics
is to embed the respective transfer matrices into a family of pairwise 
commuting operators \cite{MC}. Let  
$T_{L}\left(\omega_1^{(i)} ,\ldots,\omega_n^{(i)}  \right)$ denote the transfer matrix
on a chain of size $L$ with 
Boltzmann weights $\omega_1^{i} ,\ldots,\omega_n^{(i)} $. This approach requires that the transfer matrix
fulfill the following property,
\EQ
\left[T_{L}\left(\omega_1^{(1)},\ldots,\omega_n^{(1)} \right),
T_{L}\left(\omega_1^{(2)},\ldots,\omega_n^{(2)} \right)\right] = 0,
\label{comuteq}
\EN
for arbitrary $L$ and weights
$\omega_1^{(i)} ,\ldots,\omega_n^{(i)} $ for $i=1,2$.

At first sight it appears one needs to 
verify an infinite number of constraints for the weights
since Eq.(\ref{comuteq}) should be valid for any values of $L$. This is however not the case
since Baxter argued that
a finite number of local conditions on the weights are sufficient to assure
the commutativity among distinct transfer matrices for arbitrary $L$ \cite{BA}. These 
conditions can be
written as a matrix equation whose structure depend much on the family of lattice model under 
consideration. In what follows we shall discuss them for a relevant class of lattice systems
denominated vertex models.

Lets us for example consider a vertex model on a square lattice
of size $L \times L$. The respective statistical configurations sit on the edges of the lattice.
In the simplest case the number of states living on the horizontal and
vertical edges are the same and take values on a set of integers numbers $1,2,\ldots,N$.
Furthermore, to each vertex configuration $a,b,c,d=1,\ldots,N$ is assigned the Boltzmann weight
$\mathcal{L}\left(a,c\vert b,d\right)$ as defined in Figure 
\ref{squarelattice}.
\setlength{\unitlength}{3400sp}
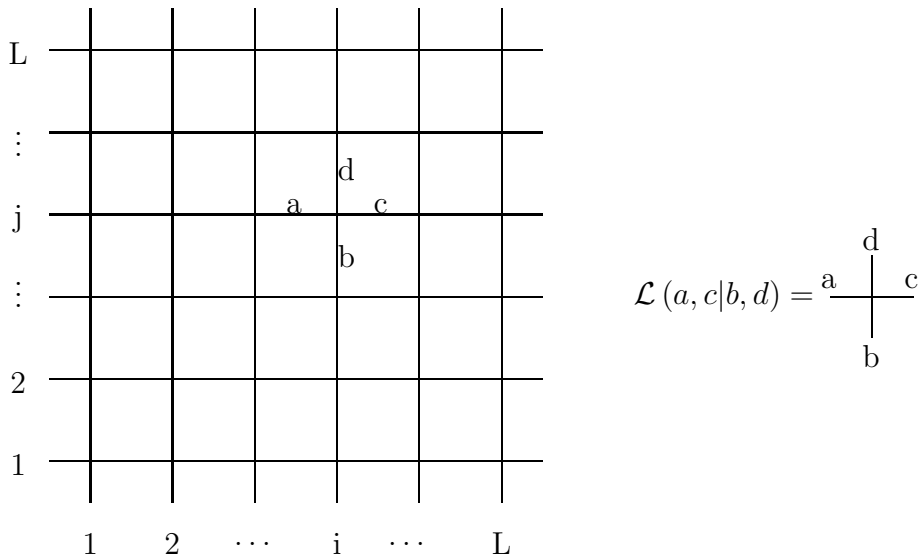
\begin{figure}[ht]
\begin{center}
\begin{picture}(6537,3912)(3376,-6361)
\put(3601,-2761){\line( 1, 0){3600}}
\put(3601,-3361){\line( 1, 0){3600}}
\put(3601,-4561){\line( 1, 0){3600}}
\put(3601,-5161){\line( 1, 0){3600}}
\put(3601,-5761){\line( 1, 0){3600}}
\put(3901,-6061){\line( 0, 1){3600}}
\put(4501,-6061){\line( 0, 1){3600}}
\put(5101,-6061){\line( 0, 1){3600}}
\put(6301,-6061){\line( 0, 1){3600}}
\put(6901,-6061){\line( 0, 1){3600}}
\put(5701,-6061){\line( 0, 1){3600}}
\put(3601,-3961){\line( 1, 0){3600}}
\put(9301,-4561){\line( 1, 0){600}}
\put(9601,-4861){\line( 0, 1){600}}
\put(3376,-5786){\makebox(0,0){1}}
\put(3376,-5186){\makebox(0,0){2}}
\put(3376,-4486){\makebox(0,0){$\vdots$}}
\put(3376,-3986){\makebox(0,0){j}}
\put(3376,-3386){\makebox(0,0){$\vdots$}}
\put(3376,-2786){\makebox(0,0){L}}
\put(3901,-6361){\makebox(0,0){1}}
\put(4501,-6361){\makebox(0,0){2}}
\put(5101,-6361){\makebox(0,0){$\dots$}}
\put(5701,-6361){\makebox(0,0){i}}
\put(6226,-6361){\makebox(0,0){$\dots$}}
\put(6901,-6361){\makebox(0,0){L}}
\put(5386,-3900){\makebox(0,0){a}}
\put(5770,-4260){\makebox(0,0){b}}
\put(6020,-3900){\makebox(0,0){c}}
\put(5770,-3631){\makebox(0,0){d}}
\put(9286,-4450){\makebox(0,0){a}}
\put(9890,-4450){\makebox(0,0){c}}
\put(9596,-4140){\makebox(0,0){d}}
\put(9596,-4990){\makebox(0,0){b}}
\put(8521,-4555){\makebox(0,0){$\mathcal{L}\left(a,c\vert b,d\right)=$}}
\end{picture} \par
\end{center}
\caption{Elementary configuration in a vertex model on a square lattice.}\label{squarelattice}
\end{figure}
The sufficient condition for the commutativity of two distinct transfer matrices associated
to a given vertex model for arbitrary $L$ is the celebrated Yang-Baxter equation \cite{BA}.
Considering the notation of Figure \ref{squarelattice},
these set of functional relations can be written as follows,
\bear
\sum_{\gamma_1 ,\gamma_2, \gamma_3=1}^{N}R\left(a_1,\ga_1\vert a_2 ,\ga_2\right)
\mathcal{L}^{(1)}\left(\ga_1,b_1\vert a_3 ,\ga_3\right)
\mathcal{L}^{(2)}\left(\ga_2,b_2\vert \ga_3 ,b_3\right) = \nonumber \\ 
\sum_{\gamma_1 ,\gamma_2, \gamma_3=1}^{N}
\mathcal{L}^{(2)}\left(a_2,\ga_2\vert a_3 ,\ga_3\right)
\mathcal{L}^{(1)}\left(a_1,\ga_1\vert \ga_3 ,b_3\right)
R\left(\ga_1,b_1\vert \ga_2 ,b_2\right).
\label{ybRLL}
\ear
where $R(a,c\vert b,d)$ are the elements of an invertible $N^2 \times N^2$ matrix
often called $R-$matrix.

In order to classify solvable vertex models with given statistical configurations
one has to find the possible solutions of the corresponding Yang-Baxter equation.
In principle, we can consider this problem by following the method devised by
Baxter for two states vertex models \cite{BA}. We start by eliminating the matrix elements $R(a,b\vert c,d)$
with the help of suitable subset of relations derived from Eq.(\ref{ybRLL}). The
remaining functional equations will then depend only on the weights $\mathcal{L}^{(1)}(a,b\vert c,d)$
and $\mathcal{L}^{(2)}(a,b\vert c,d)$
which need to be decoupled by means of the technique of separation of
variables. This step is essential to reveal us the algebraic invariants 
constraining the  Boltzmann weights parameter space.  
The parameterization of such manifolds provides
us the dependence of Boltzmann weights on spectral parameters which is for instance useful
to formulate the algebraic Bethe ansatz \cite{FAD,KO,CM}.

The implementation of all the steps described above, even for a given fixed vertex
state configuration, is in general a tantalizing problem in mathematical physics.
The concrete results are mostly concentrated on vertex models having two states
per edge, see for instance \cite{KAS,KR,WM}. In this case the non-trivial models 
that have been
uncovered are mainly associated to the algebraic manifolds of the asymmetric
six vertex model \cite{MCO}, the symmetric eight vertex model \cite{BA1} and
the so-called free-fermions systems \cite{FW}. The difficulties of the
problem increase with the number $N$ of states due to the proliferation of the 
possible allowed weights that are ultimately fixed by 
different classes of functional relations. The merit of this approach is that it
makes possible an  unambiguous classification of integrable vertex models with a given
statistical configuration from first principles.

The purpose of this paper is to begin a study of the Baxter
method for three state models satisfying the so-called ice-rule. The respective
statistical configuration leads us to the total number of
nineteen Boltzmann weights. Here we shall consider a relevant
subclass of such models whose weights are invariant by the joint
action of the parity and time ($\mathcal{PT}$) 
reversal symmetry. For this family of nineteen vertex
models we show that the dependence of the underlying algebraic
curves on the respective weights are rather universal. The possible
solvable vertex models are classified by the distinct branches of certain
invariants values entering the definition of these manifolds.
This general analysis provides us the classification of nineteen vertex models
in four different families. The first three of them
have already been discovered in the context of integrable spin-1 
\textit{XXZ} chain \cite{FZ}, the quantum inverse scattering of
the Mikhailov-Shabat model \cite{IK} and the quantum algebra
$U_q[SU(2)]$ at roots of unity \cite{AKU,COT,SI}. Interesting enough, our results reveal that these
models have the same underlying algebraic background 
despite their rather distinct quantum 
group origin.
To the best of our knowledge
the fourth uncovered vertex model is new in the literature.

We have organized this paper as follows. For sake of completeness
we review in next section the main characteristics of the nineteen vertex models.
In section \ref{secYB}  we consider the analysis of the functional relations coming from
the Yang-Baxter equation. We develop a systematic way to solve
such Yang-Baxter relations leading us to determine the algebraic curves fulfilled
by the Boltzmann weights. It turns out that the principal algebraic curve is a conic involving three
basic independent weights. The remaining amplitudes of the parameter space 
are remarkably resolved in terms
of ratios of polynomials depending on such basic weights. This makes it possible to 
classify the $\mathcal{PT}$ invariant
nineteen vertex models from a unified perspective. In section \ref{secPARA} we discuss the
parameterization of the Boltzmann weights in terms of a spectral parameter and 
the associated spin-1 Hamiltonians.
We note that the geometric form of the fundamental algebraic curve
is directly related to the nature of the excitations of the corresponding spin chains.
In section \ref{secAPP} we present the eigenvalues of the transfer matrix associated
to the new nineteen vertex model and its respective Bethe ansatz equation.
We investigate the bulk properties of this model providing additional
support to the mentioned relationship among the curve geometry and the behaviour
of the spin-1 chain excitations. Our conclusions 
are presented in section \ref{secCON}. In the appendices we
summarize some technical details  useful for the understanding of the
main text.

\section{The nineteen vertex model}
\label{nine}

In this section we shall review the main features of the nineteen vertex
model. This lattice model has three states per edge which here will be
denoted by $a,b,c,d = 0, \pm$. The 
allowed statistical configurations compatible with the
ice-rule $a+b = c+d$ lead us to nineteen different Boltzmann weights. These weights
are represented in Figure \ref{19vertex} where the respective subscripts emphasizes the non-null
charge sectors $a+b$.

\setlength{\unitlength}{2000sp}
\begin{figure}[ht]
\begin{center}
\begin{picture}(8000,8000)
{\put(100,7200){\line(1,0){1400}}}
{\put(1700,7200){\line(1,0){1400}}}
{\put(3300,7200){\line(1,0){1400}}}
{\put(4900,7200){\line(1,0){1400}}}
{\put(6500,7200){\line(1,0){1400}}}
{\put(800,7900){\line(0,-1){1400}}}
{\put(2400,7900){\line(0,-1){1400}}}
{\put(4000,7900){\line(0,-1){1400}}}
{\put(5600,7900){\line(0,-1){1400}}}
{\put(7200,7900){\line(0,-1){1400}}}
{\put(200,7350){\makebox(0,0){\fontsize{12}{14}\selectfont $+$}}}
{\put(1800,7350){\makebox(0,0){\fontsize{12}{14}\selectfont $+$}}}
{\put(3400,7350){\makebox(0,0){\fontsize{12}{14}\selectfont $0$}}}
{\put(5000,7350){\makebox(0,0){\fontsize{12}{14}\selectfont $+$}}}
{\put(6600,7350){\makebox(0,0){\fontsize{12}{14}\selectfont $0$}}}
{\put(1400,7350){\makebox(0,0){\fontsize{12}{14}\selectfont $+$}}}
{\put(3000,7350){\makebox(0,0){\fontsize{12}{14}\selectfont $+$}}}
{\put(4600,7350){\makebox(0,0){\fontsize{12}{14}\selectfont $0$}}}
{\put(6200,7350){\makebox(0,0){\fontsize{12}{14}\selectfont $0$}}}
{\put(7800,7350){\makebox(0,0){\fontsize{12}{14}\selectfont $+$}}}
{\put(950,7800){\makebox(0,0){\fontsize{12}{14}\selectfont $+$}}}
{\put(2550,7800){\makebox(0,0){\fontsize{12}{14}\selectfont $0$}}}
{\put(4150,7800){\makebox(0,0){\fontsize{12}{14}\selectfont $+$}}}
{\put(5750,7800){\makebox(0,0){\fontsize{12}{14}\selectfont $+$}}}
{\put(7350,7800){\makebox(0,0){\fontsize{12}{14}\selectfont $0$}}}
{\put(950,6600){\makebox(0,0){\fontsize{12}{14}\selectfont $+$}}}
{\put(2550,6600){\makebox(0,0){\fontsize{12}{14}\selectfont $0$}}}
{\put(4150,6600){\makebox(0,0){\fontsize{12}{14}\selectfont $+$}}}
{\put(5750,6600){\makebox(0,0){\fontsize{12}{14}\selectfont $0$}}}
{\put(7350,6600){\makebox(0,0){\fontsize{12}{14}\selectfont $+$}}}
{\put(800,6200){\makebox(0,0){\fontsize{12}{14}\selectfont $a_{+}$}}}
{\put(2400,6200){\makebox(0,0){\fontsize{12}{14}\selectfont $b_{+}$}}}
{\put(4000,6200){\makebox(0,0){\fontsize{12}{14}\selectfont $\tilde{b}_{+}$}}
}
{\put(5600,6200){\makebox(0,0){\fontsize{12}{14}\selectfont $c_{+}$}}}
{\put(7200,6200){\makebox(0,0){\fontsize{12}{14}\selectfont $\tilde{c}_{+}$}}
}

{\put(100,5100){\line(1,0){1400}}}
{\put(1700,5100){\line(1,0){1400}}}
{\put(3300,5100){\line(1,0){1400}}}
{\put(4900,5100){\line(1,0){1400}}}
{\put(6500,5100){\line(1,0){1400}}}
{\put(800,5800){\line(0,-1){1400}}}
{\put(2400,5800){\line(0,-1){1400}}}
{\put(4000,5800){\line(0,-1){1400}}}
{\put(5600,5800){\line(0,-1){1400}}}
{\put(7200,5800){\line(0,-1){1400}}}
{\put(200,5250){\makebox(0,0){\fontsize{12}{14}\selectfont $+$}}}
{\put(1800,5250){\makebox(0,0){\fontsize{12}{14}\selectfont $0$}}}
{\put(3400,5250){\makebox(0,0){\fontsize{12}{14}\selectfont $0$}}}
{\put(5000,5250){\makebox(0,0){\fontsize{12}{14}\selectfont $-$}}}
{\put(6600,5250){\makebox(0,0){\fontsize{12}{14}\selectfont $+$}}}
{\put(1400,5250){\makebox(0,0){\fontsize{12}{14}\selectfont $0$}}}
{\put(3000,5250){\makebox(0,0){\fontsize{12}{14}\selectfont $+$}}}
{\put(4600,5250){\makebox(0,0){\fontsize{12}{14}\selectfont $-$}}}
{\put(6200,5250){\makebox(0,0){\fontsize{12}{14}\selectfont $0$}}}
{\put(7800,5250){\makebox(0,0){\fontsize{12}{14}\selectfont $+$}}}
{\put(950,5700){\makebox(0,0){\fontsize{12}{14}\selectfont $0$}}}
{\put(2550,5700){\makebox(0,0){\fontsize{12}{14}\selectfont $-$}}}
{\put(4150,5700){\makebox(0,0){\fontsize{12}{14}\selectfont $+$}}}
{\put(5750,5700){\makebox(0,0){\fontsize{12}{14}\selectfont $0$}}}
{\put(7350,5700){\makebox(0,0){\fontsize{12}{14}\selectfont $-$}}}
{\put(950,4500){\makebox(0,0){\fontsize{12}{14}\selectfont $-$}}}
{\put(2550,4500){\makebox(0,0){\fontsize{12}{14}\selectfont $0$}}}
{\put(4150,4500){\makebox(0,0){\fontsize{12}{14}\selectfont $0$}}}
{\put(5750,4500){\makebox(0,0){\fontsize{12}{14}\selectfont $+$}}}
{\put(7350,4500){\makebox(0,0){\fontsize{12}{14}\selectfont $-$}}}
{\put(800,4100){\makebox(0,0){\fontsize{12}{14}\selectfont $d$}}}
{\put(2400,4100){\makebox(0,0){\fontsize{12}{14}\selectfont $\tilde{d}$}}
}
{\put(4000,4100){\makebox(0,0){\fontsize{12}{14}\selectfont $e$}}
}
{\put(5600,4100){\makebox(0,0){\fontsize{12}{14}\selectfont $\tilde{e}$}}
}
{\put(7200,4100){\makebox(0,0){\fontsize{12}{14}\selectfont $f$}}
}
{\put(100,3000){\line(1,0){1400}}}
{\put(1700,3000){\line(1,0){1400}}}
{\put(3300,3000){\line(1,0){1400}}}
{\put(4900,3000){\line(1,0){1400}}}
{\put(6500,3000){\line(1,0){1400}}}
{\put(800,3700){\line(0,-1){1400}}}
{\put(2400,3700){\line(0,-1){1400}}}
{\put(4000,3700){\line(0,-1){1400}}}
{\put(5600,3700){\line(0,-1){1400}}}
{\put(7200,3700){\line(0,-1){1400}}}
{\put(200,3150){\makebox(0,0){\fontsize{12}{14}\selectfont $-$}}}
{\put(1800,3150){\makebox(0,0){\fontsize{12}{14}\selectfont $0$}}}
{\put(3400,3150){\makebox(0,0){\fontsize{12}{14}\selectfont $+$}}}
{\put(5000,3150){\makebox(0,0){\fontsize{12}{14}\selectfont $-$}}}
{\put(6600,3150){\makebox(0,0){\fontsize{12}{14}\selectfont $-$}}}
{\put(1400,3150){\makebox(0,0){\fontsize{12}{14}\selectfont $-$}}}
{\put(3000,3150){\makebox(0,0){\fontsize{12}{14}\selectfont $0$}}}
{\put(4600,3150){\makebox(0,0){\fontsize{12}{14}\selectfont $-$}}}
{\put(6200,3150){\makebox(0,0){\fontsize{12}{14}\selectfont $+$}}}
{\put(7800,3150){\makebox(0,0){\fontsize{12}{14}\selectfont $-$}}}
{\put(950,3600){\makebox(0,0){\fontsize{12}{14}\selectfont $+$}}}
{\put(2550,3600){\makebox(0,0){\fontsize{12}{14}\selectfont $0$}}}
{\put(4150,3600){\makebox(0,0){\fontsize{12}{14}\selectfont $+$}}}
{\put(5750,3600){\makebox(0,0){\fontsize{12}{14}\selectfont $-$}}}
{\put(7350,3600){\makebox(0,0){\fontsize{12}{14}\selectfont $-$}}}
{\put(950,2400){\makebox(0,0){\fontsize{12}{14}\selectfont $+$}}}
{\put(2550,2400){\makebox(0,0){\fontsize{12}{14}\selectfont $0$}}}
{\put(4150,2400){\makebox(0,0){\fontsize{12}{14}\selectfont $-$}}}
{\put(5750,2400){\makebox(0,0){\fontsize{12}{14}\selectfont $+$}}}
{\put(7350,2400){\makebox(0,0){\fontsize{12}{14}\selectfont $-$}}}
{\put(800,2000){\makebox(0,0){\fontsize{12}{14}\selectfont $\tilde{f}$}}}
{\put(2400,2000){\makebox(0,0){\fontsize{12}{14}\selectfont $g$}}
}
{\put(4000,2000){\makebox(0,0){\fontsize{12}{14}\selectfont $h$}}
}
{\put(5600,2000){\makebox(0,0){\fontsize{12}{14}\selectfont $\tilde{h}$}}
}
{\put(7200,2000){\makebox(0,0){\fontsize{12}{14}\selectfont $a_{-}$}}
}
{\put(900,900){\line(1,0){1400}}}
{\put(2500,900){\line(1,0){1400}}}
{\put(4100,900){\line(1,0){1400}}}
{\put(5800,900){\line(1,0){1400}}}
{\put(1600,1600){\line(0,-1){1400}}}
{\put(3200,1600){\line(0,-1){1400}}}
{\put(4800,1600){\line(0,-1){1400}}}
{\put(6500,1600){\line(0,-1){1400}}}
{\put(1000,1050){\makebox(0,0){\fontsize{12}{14}\selectfont $-$}}}
{\put(2600,1050){\makebox(0,0){\fontsize{12}{14}\selectfont $0$}}}
{\put(4200,1050){\makebox(0,0){\fontsize{12}{14}\selectfont $0$}}}
{\put(5900,1050){\makebox(0,0){\fontsize{12}{14}\selectfont $-$}}}
{\put(2200,1050){\makebox(0,0){\fontsize{12}{14}\selectfont $-$}}}
{\put(3800,1050){\makebox(0,0){\fontsize{12}{14}\selectfont $0$}}}
{\put(5400,1050){\makebox(0,0){\fontsize{12}{14}\selectfont $-$}}}
{\put(7100,1050){\makebox(0,0){\fontsize{12}{14}\selectfont $0$}}}
{\put(1750,1500){\makebox(0,0){\fontsize{12}{14}\selectfont $0$}}}
{\put(3350,1500){\makebox(0,0){\fontsize{12}{14}\selectfont $-$}}}
{\put(4950,1500){\makebox(0,0){\fontsize{12}{14}\selectfont $0$}}}
{\put(6650,1500){\makebox(0,0){\fontsize{12}{14}\selectfont $-$}}}
{\put(1750,300){\makebox(0,0){\fontsize{12}{14}\selectfont $0$}}}
{\put(3350,300){\makebox(0,0){\fontsize{12}{14}\selectfont $-$}}}
{\put(4950,300){\makebox(0,0){\fontsize{12}{14}\selectfont $-$}}}
{\put(6650,300){\makebox(0,0){\fontsize{12}{14}\selectfont $0$}}}
{\put(1700,-100){\makebox(0,0){\fontsize{12}{14}\selectfont $b_{-}$}}}
{\put(3300,-100){\makebox(0,0){\fontsize{12}{14}\selectfont $\tilde{b}_{-}$}}}
{\put(4900,-100){\makebox(0,0){\fontsize{12}{14}\selectfont $c_{-}$}}}
{\put(6500,-100){\makebox(0,0){\fontsize{12}{14}\selectfont $\tilde{c}_{-}$}}}
\end{picture}
\end{center}
\caption{The vertex configurations of a general nineteen vertex model
on a square lattice.}\label{19vertex}
\end{figure}
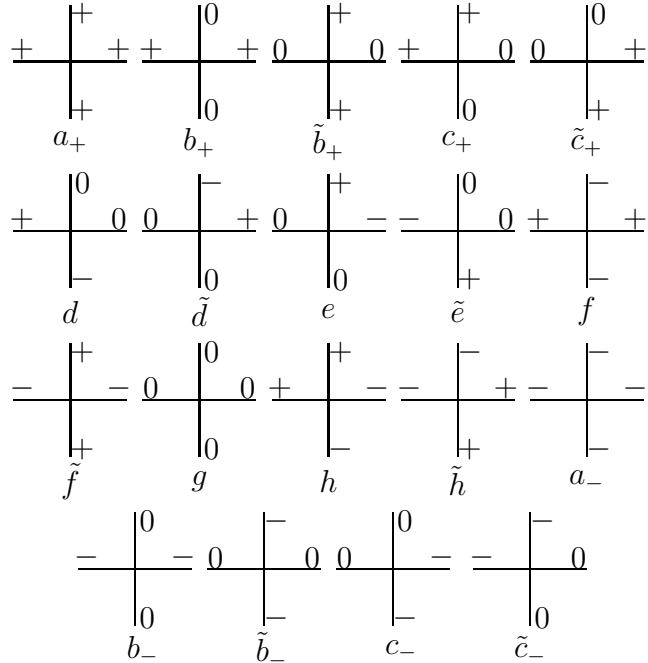
In this paper we will investigate the integrable manifolds of nineteen
vertex models whose weights are $\mathcal{PT}$ invariant.
This invariance relates some of the weights leading
us to subclass of three-state vertex models with fourteen distinct Boltzmann
weights. More precisely, the $\mathcal{PT}$ symmetry imposes 
the following relationship
among the weights of Figure \ref{19vertex},
\EQ
\tilde{b}_{\pm} = b_{\pm},\, \tilde{f} = f,\, e = d,\, \tilde{e} = \tilde{d}.
\label{PTsymmetryOnWeights}
\EN

We begin by introducing a common notation for 
the Boltzmann weights of such family of
nineteen vertex models. Taking into account the parameter subspace
(\ref{PTsymmetryOnWeights}) one can represent the content
of the $\mathcal{L}^{(i)}$-operators  by the following matrix,
\begin{eqnarray}
\mathcal{L}^{(i)} & =& a_{+}^{(i)} e_{1,1} \otimes e_{1,1}
+b_{+}^{(i)} \left [ e_{1,1} \otimes e_{2,2} +
e_{2,2} \otimes e_{1,1} \right ]
+f^{(i)} \left [ e_{1,1} \otimes e_{3,3} +
e_{3,3} \otimes e_{1,1} \right ]
\nonumber \\
&+&b_{-}^{(i)} \left [ e_{2,2} \otimes e_{3,3} 
+ e_{3,3} \otimes e_{2,2} \right ]
+g^{(i)} e_{2,2} \otimes e_{2,2}
+ a_{-}^{(i)} e_{3,3} \otimes e_{3,3}
+ h^{(i)} e_{1,3} \otimes e_{3,1}
\nonumber \\
&+& \tilde{h}^{(i)} e_{3,1} \otimes e_{1,3}
+c_{+}^{(i)} e_{1,2} \otimes e_{2,1} +
\tilde{c}_{+}^{(i)} e_{2,1} \otimes e_{1,2} 
+c_{-}^{(i)} e_{2,3} \otimes e_{3,2} +
\tilde{c}_{-}^{(i)} e_{3,2} \otimes e_{2,3} 
\nonumber \\
&+&d^{(i)} \left [ e_{1,2} \otimes e_{3,2} 
+ e_{2,3} \otimes e_{2,1} \right ]
+\tilde{d}^{(i)} \left [ e_{2,1} \otimes e_{2,3} 
+ e_{3,2} \otimes e_{1,2} \right ],~~\mathrm{for}~~i=1,2,
\label{Lmatrix}
\end{eqnarray}
where $e_{a,b}$ denote $3 \times 3$ Weyl matrices.

By substituting the $\mathcal{L}^{(i)}-$operators (\ref{Lmatrix}) in the Yang-Baxter
equation (\ref{ybRLL}) we are able to determine the null matrix elements of the $R-$matrix.
Under the mild assumption that the weight $a_{\pm}^{(i)}$
are in general distinct from  
$b_{\pm}^{(i)}$ and 
$f^{(i)}$  
it is not difficult to see 
that the $R-$matrix has the same  form
of the $\mathcal{L}^{(i)}-$operators. For latter convenience we shall therefore express the  
$R-$matrix as,
\begin{eqnarray}
R & =& a_{+}^{(0)} e_{1,1} \otimes e_{1,1}
+b_{+}^{(0)} \left [ e_{1,1} \otimes e_{2,2} +
e_{2,2} \otimes e_{1,1} \right ]
+f^{(0)} \left [ e_{1,1} \otimes e_{3,3} +
e_{3,3} \otimes e_{1,1} \right ]
\nonumber \\
&+&b_{-}^{(0)} \left [ e_{2,2} \otimes e_{3,3} 
+ e_{3,3} \otimes e_{2,2} \right ]
+g^{(0)} e_{2,2} \otimes e_{2,2}
+ a_{-}^{(0)} e_{3,3} \otimes e_{3,3}
+ h^{(0)} e_{1,3} \otimes e_{3,1}
\nonumber \\
&+& \tilde{h}^{(0)} e_{3,1} \otimes e_{1,3}
+c_{+}^{(0)} e_{1,2} \otimes e_{2,1} +
\tilde{c}_{+}^{(0)} e_{2,1} \otimes e_{1,2} 
+c_{-}^{(0)} e_{2,3} \otimes e_{3,2} +
\tilde{c}_{-}^{(0)} e_{3,2} \otimes e_{2,3} 
\nonumber \\
&+&d^{(0)} \left [ e_{1,2} \otimes e_{3,2} 
+ e_{2,3} \otimes e_{2,1} \right ]
+\tilde{d}^{(0)} \left [ e_{2,1} \otimes e_{2,3} 
+ e_{3,2} \otimes e_{1,2} \right ] .
\label{Rmatrix}
\end{eqnarray}

We emphasize here that we are interested to classify genuine nineteen vertex models and therefore
all the weights 
$a_{\pm}^{(i)}$,
$b_{\pm}^{(i)}$,
$c_{\pm}^{(i)}$,
$\tilde{c}_{\pm}^{(i)}$,
$d^{(i)}$,
$\tilde{d}^{(i)}$,
$f^{(i)}$, $g^{(i)}$,
$h^{(i)}$ and
$\tilde{h}^{(i)}$ for $i=0,1,2$  are assumed to be non-null. We recall that a classification
of solvable three-state vertex models has been attempted before in the literature \cite{JAPA}.
In the work \cite{JAPA}, more 
stringent symmetry conditions for the weights were assumed besides the fact that some of them could
be null. As a result,  
the only strict nineteen vertex model found was the standard
Fateev-Zamolodchikov spin one model \cite{FZ}.

\section{The Yang-Baxter Equation}
\label{secYB}
The purpose of this section is to investigate the functional relations for the Boltzmann
weights which are derived by substituting the $\mathcal{L}^{(i)}-$operator (\ref{Lmatrix}) and the
$R$-matrix (\ref{Rmatrix}) structures into the Yang-Baxter equation (\ref{ybRLL}). We
find that the functional relations can be classified in terms of the number of 
distinct triple product
of weights. It turns out the minimum number of triple products is two while the
maximum is five. In Table \ref{tab1} we summarize the number of different
equations having two, three, four and five types of triple product of weights. Clearly,
both the number and the structure of the functional 
relations to be analyzed are far more involving than that associated to the
$N = 2$ state vertex model satisfying the ice-rule \footnote{In this case we have an asymmetric
six-vertex model and we end up with six non-trivial relations containing 
only three triple products. Their solution is easily found 
by imposing that two distinct $3 \times 3$
determinants made out of the six relations are null \cite{KAS}.}. We shall therefore 
start our discussion by first
solving the simplest relations containing two triple products.

\begin{table}[ht]
\begin{center}
\begin{tabular}{|c|c|}
\hline
Number of triple products & Number of equations \\ \hline
two & 6 \\ \hline
three & 36 \\ \hline
four & 57 \\ \hline
five & 24 \\ \hline
\end{tabular}
\caption{The dependence of the number of distinct functional relations on
the respective number of triple products.}
\label{tab1}
\end{center}
\end{table}

\subsection{Two terms relations}
The six equations possessing only two different types of triple product of weights are given by,
\EQ \label{eq9e84}
c_{\pm}^{(0)}\tilde{c}^{(1)}_{\pm}c_{\pm}^{(2)} - 
\tilde{c}_{\pm}^{(0)}c_{\pm}^{(1)}\tilde{c}_{\pm}^{(2)} = 0,
\EN 
\EQ \label{eq15e76}
d^{(0)}\tilde{d}^{(1)}c_{\pm}^{(2)} - 
\tilde{d}^{(0)}d^{(1)}\tilde{c}_{\pm}^{(2)} = 0,
\EN
\EQ \label{eq33e57}
c_{\pm}^{(0)}\tilde{d}^{(1)}d^{(2)} - \tilde{c}_{\pm}^{(0)}d^{(1)}\tilde{d}^{(2)} = 0.
\EN
We first note that the
apparent difference between $ c^{(i)}_{\pm} $ and $ \tilde{c}^{(i)}_{\pm} $ can be gauge away by
a transformation preserving the Yang-Baxter equation. Without losing generality
we can set,
\EQ\label{gaugetransf}
\tilde{c}_{\pm}^{(i)} = c_{\pm}^{(i)},~~ \mathrm{for}~~i=0,1,2.
\EN
By substituting the result (\ref{gaugetransf}) in
Eqs.(\ref{eq15e76},\ref{eq33e57}) 
we conclude that the weight 
$\tilde{d}^{(i)}$ becomes proportional to $d^{(i)}$, 
\EQ\label{invPSI}
\frac{\tilde{d}^{(i)}}{d^{(i)}} = \Psi,~~ \mathrm{for}~~i=0,1,2,
\EN
where $ \Psi $ is our first invariant value.

We shall now turn our attention to the relations involving three types of
triple products.

\subsection{Three terms relations}
By using the solution (\ref{gaugetransf},\ref{invPSI}) we find that the thirty six relations
possessing three triple products are pairwise equivalent leading us to eighteen distinct
functional equations. They can be classified in terms of three different groups of
six equations given by,

\begin{itemize}
\item group $G_{\pm}^{(1)}$
\EQ \label{eq2e120}
b_{\pm}^{(0)}c_{\pm}^{(1)}c_{\pm}^{(2)} + c_{\pm}^{(0)}a_{\pm}^{(1)}b_{\pm}^{(2)}
- c_{\pm}^{(0)}b_{\pm}^{(1)}a_{\pm}^{(2)} = 0, 
\EN
\EQ \label{eq1e88}
b_{\pm}^{(0)}c_{\pm}^{(1)}b_{\pm}^{(2)} + c_{\pm}^{(0)}a_{\pm}^{(1)}c_{\pm}^{(2)} 
- a_{\pm}^{(0)}c_{\pm}^{(1)}a_{\pm}^{(2)} = 0, 
\EN
\EQ \label{eq8e89}
c_{\pm}^{(0)}c_{\pm}^{(1)}b_{\pm}^{(2)} + b_{\pm}^{(0)}a_{\pm}^{(1)}c_{\pm}^{(2)} 
- a_{\pm}^{(0)}b_{\pm}^{(1)}c_{\pm}^{(2)} = 0, 
\EN

\item group $G_{\pm}^{(2)}$
\EQ \label{eq7e114}
d^{(0)}b_{\pm}^{(1)}b_{\pm}^{(2)} + f^{(0)}d^{(1)}c_{\pm}^{(2)} 
- d^{(0)}f^{(1)}a_{\pm}^{(2)} = 0, 
\EN
\EQ \label{eq6e82}
f^{(0)}d^{(1)}b_{\pm}^{(2)} + d^{(0)}b_{\pm}^{(1)}c_{\pm}^{(2)} 
- b_{\pm}^{(0)}d^{(1)}a_{\pm}^{(2)}  = 0, 
\EN
\EQ \label{eq22e75}
\Psi d^{(0)}d^{(1)}b_{\mp}^{(2)} + f^{(0)}b_{\pm}^{(1)}c_{\mp}^{(2)} 
- b_{\pm}^{(0)}f^{(1)}c_{\mp}^{(2)} = 0, 
\EN

\item group $G_{\pm}^{(3)}$
\EQ \label{eq56e40}
b_{\pm}^{(0)}b_{\pm}^{(1)}d^{(2)} + c_{\pm}^{(0)}d^{(1)}f^{(2)} 
- a_{\pm}^{(0)}f^{(1)}d^{(2)} = 0, 
\EN
\EQ \label{eq43e38}
c_{\pm}^{(0)}b_{\pm}^{(1)}d^{(2)} + b_{\pm}^{(0)}d^{(1)}f^{(2)} 
- a_{\pm}^{(0)}d^{(1)}b_{\pm}^{(2)} = 0, 
\EN
\EQ \label{eq95e21}
\Psi b_{\pm}^{(0)}d^{(1)}d^{(2)} + c_{\pm}^{(0)}b_{\mp}^{(1)}f^{(2)} 
- c_{\pm}^{(0)}f^{(1)}b_{\mp}^{(2)} = 0 ,
\EN
\end{itemize}
where the subscript $\pm$ in Eqs.(\ref{eq2e120}-\ref{eq95e21}) means that each of them
splits into
two independent functional relations.

Altogether we have eighteen linear homogeneous 
equations but only eight Boltzmann
weights $ a_{\pm}^{(0)} $,$ b_{\pm}^{(0)} $,$ c_{\pm}^{(0)} $,
$ d^{(0)} $ and $ f^{(0)} $ are at our
disposal to be eliminated. Therefore, we have a high degree of over-determination
to overcome making the solution of Eqs.(\ref{eq2e120}-\ref{eq95e21}) far from being trivial.
We shall start
our analysis by considering the group $G_{\pm}^{(1)}$ of equations. They are similar
to the functional equations underlying the symmetric six-vertex model and can be  easily handled. We
first eliminate the weights $ a_{\pm}^{(0)} $, $ b_{\pm}^{(0)} $ 
in terms of $ c_{\pm}^{(0)} $ with the
help of Eqs.(\ref{eq2e120},\ref{eq1e88}). As a result we obtain,
\EQ \label{expv0apv0am}
\frac{a_{\pm}^{(0)}}{c_{\pm}^{(0)}} = 
\frac{b_{\pm}^{(1)}a_{\pm}^{(2)}b_{\pm}^{(2)}-a_{\pm}^{(1)}
\left \{\left[b_{\pm}^{(2)}\right]^2-\left[c_{\pm}^{(2)}\right]^2
\right \} }
{c_{\pm}^{(1)}a_{\pm}^{(2)}c_{\pm}^{(2)}},
\EN
\EQ \label{expv0bpv0bm}
\frac{b_{\pm}^{(0)}}{c_{\pm}^{(0)}} = \frac{b_{\pm}^{(1)}a_{\pm}^{(2)} 
- a_{\pm}^{(1)}b_{\pm}^{(2)}}{c_{\pm}^{(1)}c_{\pm}^{(2)}}.
\EN
By substituting the above results 
(\ref{expv0apv0am},\ref{expv0bpv0bm}) in Eq.(\ref{eq8e89}) 
we find two
separable curves that involve 
the weights $ a_{\pm}^{(i)}$,$b_{\pm}^{(i)}$ and $c_{\pm}^{(i)} $ for $i=1,2$. 
Their solution leads us to invariants typical of six-vertex models, 
\EQ\label{invDELTAPM}
\frac{\left[a_{\pm}^{(i)}\right]^{2} + \left[b_{\pm}^{(i)}\right]^{2} - \left[c_{\pm}^{(i)}\right]^{2}}
{a_{\pm}^{(i)}b_{\pm}^{(i)}} 
= \Delta_{\pm},~~\mathrm{for}~~~i=1,2,
\EN
where $ \Delta_{\pm} $ are free invariant parameters.

We now turn our attention to the relations associated to the group $G_{\pm}^{(2)}$. 
We start by first eliminating
the weights $d^{(0)}$ and $f^{(0)}$ from Eqs.(\ref{eq7e114},\ref{eq6e82}). Because 
they can be isolated
from equations
possessing different charge sectors we need to keep track of their  explicit expressions.
The compatibility of such distinct solutions will be implemented subsequently. The expressions 
for $d^{(0)}$ and $f^{(0)}$
are,
\EQ \label{expv0d}
\frac{d^{(0)}}{c_{\pm}^{(0)}} = \frac{d^{(1)}a_{\pm}^{(2)}}{c_{\pm}^{(1)}}
\left\{\frac{b_{\pm}^{(1)}a_{\pm}^{(2)} - a_{\pm}^{(1)}b_{\pm}^{(2)}}
{f^{(1)}a_{\pm}^{(2)}b_{\pm}^{(2)} - b_{\pm}^{(1)}\left[b_{\pm}^{(2)}\right]^2 
+ b_{\pm}^{(1)} \left[c_{\pm}^{(2)}\right]^2}\right\},
\EN
\EQ \label{expv0f}
f^{(0)} = \left[\frac{f^{(1)}a_{\pm}^{(2)} - b_{\pm}^{(1)}b_{\pm}^{(2)}}{d^{(1)}c_{\pm}^{(2)}}\right]d^{(0)},
\EN
where as before they are given in terms of the common weights $c_{\pm}^{(0)}$.

We then substitute Eqs.(\ref{expv0d},\ref{expv0f}) in Eq.(\ref{eq22e75}) and by 
taking into account
that the Boltzmann weights are non-null we find the following relation,
\EQ \label{expdetG2}
c_{\mp}^{(2)}\left\{\left(\left[b_{\pm}^{(1)}\right]^{2}+\left[f^{(1)}\right]^{2}\right)
a_{\pm}^{(2)}b_{\pm}^{(2)}-b_{\pm}^{(1)}f^{(1)}\left(\left[a_{\pm}^{(2)}\right]^{2}
+ \left[b_{\pm}^{(2)}\right]^{2} - \left[c_{\pm}^{(2)}\right]^{2}\right)\right\}
-\Psi \left[d^{(1)}\right]^2 a_{\pm}^{(2)}b_{\mp}^{(2)}c_{\pm}^{(2)} = 0.
\EN
By using the previous invariant (\ref{invDELTAPM}) on the second term of Eq.(\ref{expdetG2}) 
one is able to split
the weights with index $i=1$ from those labeled by $i=2$.
The solution of Eq.(\ref{expdetG2}) leads us to
the new constraints,
\EQ \label{invLAPM}
\frac{\left[b_{\pm}^{(1)}\right]^2+\left[f^{(1)}\right]^2-\Delta_{\pm}b_{\pm}^{(1)}f^{(1)}}{\left[d^{(1)}\right]^2} = 
\Psi\frac{b_{\mp}^{(2)}c_{\pm}^{(2)}}{b_{\pm}^{(2)}c_{\mp}^{(2)}} =
\Lambda_{\pm},
\EN
where $ \Lambda_{\pm} $ are once again constant parameters. However, due to the consistency
of the right hand side of Eq.(\ref{invLAPM}) they are related by,
\EQ \label{constraintLAM}
\LAP \LAM = \Psi^2.
\EN
In order to complete the analysis of the group $G_{\pm}^{(2)}$ we still need to impose
the compatibility between the two possibilities for $d^{(0)}$ and $f^{(0)}$ 
derived from relations
(\ref{expv0d},\ref{expv0f}). The consistency for the weight $d^{(0)}$ is easily resolved
by fixing
the relation between the amplitudes $c_+^{(0)}$ and $c_-^{(0)}$. By way of contrast 
the compatibility
for weight $f^{(0)}$ requires us to identify the expressions for
the charge indices $\pm$ of Eq.(\ref{expv0f}). The result of such identification is,
\EQ \label{expcompv0f}
b_{-}^{(1)}b_{-}^{(2)}c_{+}^{(2)} - b_{+}^{(1)}b_{+}^{(2)}c_{-}^{(2)} 
+ f^{(1)}a_{+}^{(2)}c_{-}^{(2)} - f^{(1)}a_{-}^{(2)}c_{+}^{(2)} = 0.
\EN
Fortunately, Eq.(\ref{expcompv0f}) becomes separable once we 
take into account the right hand side of
Eq.(\ref{invLAPM}). This allows us to relate the products 
$b_{-}^{(2)}c_{+}^{(2)}$ with $b_{+}^{(2)}c_{-}^{(2)}$
and as a result Eq.(\ref{expcompv0f}) can be solved by the method of separation of variables. 
The solution is,
\EQ \label{invOMEGA}
\frac{\frac{\Lambda_{+}}{\Psi}b_{-}^{(1)}-b_{+}^{(1)}}{f^{(1)}} =
\frac{a_{-}^{(2)}c_{+}^{(2)}-a_{+}^{(2)}c_{-}^{(2)}}{b_{+}^{(2)}c_{-}^{(2)}} = \Omega,
\EN
where $\Omega$ is a new invariant value.

Before proceeding with the analysis of group $G_{\pm}^{(3)}$
we should pause to discuss the results obtained so far. The
main feature of the last invariant (\ref{invOMEGA}) is
that it relates weights with different charge index. We
expect therefore that the invariants values 
$\dpm$, $\LAP$,$\Psi$ and $\OM$ should not be independent
of each other. One way to unveil such constraints is to proceed
as follows.
By using the left hand side of Eqs.(\ref{invLAPM},\ref{invOMEGA})
we can eliminate the weights $d^{(1)}$ and $b_{-}^{(1)}$,
namely
\EQ\label{qv1dpEv1bm}
\left[d^{(1)}\right]^2 = \frac{1}{\LAP}\left\{\left[b_{+}^{(1)}\right]^2+
\left[f^{(1)}\right]^2-\dpp b_{+}^{(1)}f^{(1)}\right\},~~
b_{-}^{(1)} = \frac{\Psi}{\LAP}\left[b_{+}^{(1)}+\OM f^{(1)}\right].
\EN
We now substitute the results (\ref{qv1dpEv1bm}) on the left
hand side of Eq.(\ref{invLAPM}) with the charge index $-$. After
using the relation (\ref{constraintLAM}) we obtain the expression,
\EQ\label{eqvinc0poli}
\Psi\left(2\OM \Psi + \dpp\Psi - \dmm\LAP\right) b_{+}^{(1)}+
\left(\LAP^2-\dmm\LAP\OM \Psi - \Psi^2 + \OM^2 \Psi^2 \right) f^{(1)} = 0.
\EN

We next consider analogous approach to the weights
having index $i=2$. We first eliminate the weights
$b_{-}^{(2)}$ and $c_{-}^{(2)}$ with the help of the right
hand side of Eqs.(\ref{invLAPM},\ref{invOMEGA}),
\EQ\label{expv2bmv2cm}
b_{-}^{(2)}= \frac{\LAP}{\Psi}\frac{b_{+}^{(2)}c_{-}^{(2)}}{c_{+}^{(2)}},~~
c_{-}^{(2)}= \frac{a_{-}^{(2)}c_{+}^{(2)}}{a_{+}^{(2)}+\OM b_{+}^{(2)}}.
\EN
By substituting
Eq.(\ref{expv2bmv2cm}) on the invariant connecting the weights
$a_{-}^{(2)}$, $b_{-}^{(2)}$ and $c_{-}^{(2)}$ given by Eq.(\ref{invDELTAPM})
one easily obtains,
\EQ\label{eqcomp1}
\LAP^2\left[b_{+}^{(2)}\right]^2 - \dmm\LAP\Psi b_{+}^{(2)}\left[a_{+}^{(2)} + \OM b_{+}^{(2)}\right] +
\Psi^2\left\{\left[a_{+}^{(2)}\right]^2 + 2\OM a_{+}^{(2)}b_{+}^{(2)}+
\OM^2 \left[b_{+}^{(2)}\right]^2 - \left[c_{+}^{(2)}\right]^2\right\} = 0.
\EN
The above relation can be further simplified by using in the
last term of Eq.(\ref{eqcomp1}) the expression for
the invariant $\dpp$, see Eq.(\ref{invDELTAPM}). By performing such simplification we find,
\EQ\label{eqvinc1poli}
\Psi\left(2\OM \Psi + \dpp\Psi - \dmm\LAP\right) a_{+}^{(2)}+
\left(\LAP^2-\dmm\LAP\OM \Psi - \Psi^2 + \OM^2 \Psi^2\right) b_{+}^{(2)} = 0.
\EN

We now reached a point in which two different possibilities emerged.
The first one consists in the assumption that the pair of weights 
$\{b_{+}^{(i)},f^{(i)}\}$ and $\{a_{+}^{(i)},b_{+}^{(i)}\}$ are considered
linearly dependent. From 
Eqs.(\ref{eqvinc0poli},\ref{eqvinc1poli})
we see that this latter  hypothesis
implies in the identity $\frac{b_{+}^{(1)}}{f^{(1)}}=\frac{a_{+}^{(2)}}{b_{+}^{(2)}}$
which together with Eq.(\ref{expv0f})
lead us to conclude that the weight $f^{(0)}$ vanishes. Here we stress that we are looking
for genuine nineteen vertex models and therefore such solution is disregarded.  
Thus, we are
left with the second possibility which is simply to set
the coefficients of Eqs.(\ref{eqvinc0poli},\ref{eqvinc1poli})
to zero. This condition implies that the invariants $\dpm$, $\LAP$, $\Psi$
and $\OM$ are constrained by the following equations,
\EQ\label{eqvinc1k1}
2\OM \Psi + \dpp\Psi - \dmm\LAP = 0,
\EN
\EQ\label{eqvinc1k2}
\LAP^2-\dmm\LAP\OM \Psi - \Psi^2 + \OM^2 \Psi^2 = 0.
\EN

Let us now discuss the solution of the functional equations associated to the group
$G_{\pm}^{(3)}$.  For this group we see that the weights at our
disposal to be eliminated are 
$a_{\pm}^{(0)}$, $b_{\pm}^{(0)}$ and $c_{\pm}^{(0)}$. However, they have   
already been 
computed by means of Eqs.(\ref{eq2e120}-\ref{eq8e89}) and therefore  our task consists to
make the equations of group 
$G_{\pm}^{(3)}$ compatible with those of group $G_{\pm}^{(1)}$. 
The first step to solve this problem is to assure from the start 
that all the six functional relations of group
$G_{\pm}^{(3)}$ are indeed satisfied. This is done by 
eliminating
the weights $a_{\pm}^{(0)}$ and $b_{\pm}^{(0)}$ from Eqs.(\ref{eq56e40},\ref{eq43e38}), 
\EQ\label{expv0apv0amG3}
\frac{a_{\pm}^{(0)}}{c_{\pm}^{(0)}} = \frac{1}{d^{(1)}d^{(2)}}
\left\{\frac{\left[b_{\pm}^{(1)}d^{(2)}-d^{(1)}f^{(2)}\right]
\left[b_{\pm}^{(1)}d^{(2)}+d^{(1)}f^{(2)}\right]}{b_{\pm}^{(1)}b_{\pm}^{(2)}-f^{(1)}f^{(2)}}\right\},
\EN
\EQ\label{expv0bpv0bmG3}
\frac{b_{\pm}^{(0)}}{c_{\pm}^{(0)}} = \frac{1}{d^{(1)}d^{(2)}}
\left\{\frac{b_{\pm}^{(1)}f^{(1)}\left[d^{(2)}\right]^2-
\left[d^{(1)}\right]^2b_{\pm}^{(2)}f^{(2)}}{b_{\pm}^{(1)}b_{\pm}^{(2)}-f^{(1)}f^{(2)}}\right\}.
\EN

By substituting the expressions
(\ref{expv0apv0amG3},\ref{expv0bpv0bmG3}) in the last equation of the group
$G_{\pm}^{(3)}$, i.e. 
Eq.(\ref{eq95e21}),  we obtain for the charge $+$ sector,
\EQ\label{reseq95}
b_{+}^{(1)}f^{(1)}\left\{b_{-}^{(2)}b_{+}^{(2)}-\Psi \left[d^{(2)}\right]^2\right\} 
- b_{+}^{(2)}f^{(2)}\left\{b_{-}^{(1)}b_{+}^{(1)}-\Psi \left[d^{(1)}\right]^2\right\}
- \left[f^{(1)}\right]^2 b_{-}^{(2)}f^{(2)} + b_{-}^{(1)}f^{(1)}\left[f^{(2)}\right]^2 = 0,
\EN
while for the charge $-$ sector one finds,
\EQ\label{reseq21}
b_{-}^{(1)}f^{(1)}\left\{b_{-}^{(2)}b_{+}^{(2)}-\Psi \left[d^{(2)}\right]^2\right\} 
- b_{-}^{(2)}f^{(2)}\left\{b_{-}^{(1)}b_{+}^{(1)}-\Psi \left[d^{(1)}\right]^2\right\} 
- \left[f^{(1)}\right]^2 b_{+}^{(2)}f^{(2)} + b_{+}^{(1)}f^{(1)}\left[f^{(2)}\right]^2 = 0,
\EN
While 
Eqs.(\ref{reseq95},\ref{reseq21}) are not individually separable it turns out that suitable
linear combinations of such equations can be split on the indices $i=1,2$.
In fact, 
by adding and subtracting Eqs.(\ref{reseq95},\ref{reseq21}) we found that they
are solved provided the following constraint is verified,
\EQ\label{invTHETAPM}
\frac{\left[b_{+}^{(i)}\pm b_{-}^{(i)}\right]f^{(i)}}
{b_{+}^{(i)}b_{-}^{(i)}-\Psi \left[d^{(i)}\right]^2 \pm \left[f^{(i)}\right]^2} =
\Theta_{\pm},~~\mathrm{for}~~~i=1,2,
\EN
where $\TPM$ are additional invariants. 

Once again the invariant (\ref{invTHETAPM}) connects weights
carrying distinct charge sectors $\pm$. We have therefore
to repeat the same analysis we did for the previous
invariant $\OM$, see Eq.(\ref{invOMEGA}). As before,
the expression (\ref{invTHETAPM}) for index $i=1$
is simplified with the help of the weights $d^{(1)}$
and $b_{-}^{(1)}$ given by Eq.(\ref{qv1dpEv1bm}). This leads us
to the following polynomial identity,
\EQ\label{eqvinc2poli}
\left[\LAP \pm \Psi - \Psi\TPM(\dpp+\OM)\right] b_{+}^{(1)} +
\left[\pm \OM\Psi \mp \TPM(\LAP\mp \Psi)\right] f^{(1)} = 0.
\EN
Considering that we are looking for linearly independent
solutions for the weights $b_{+}^{(1)}$ and $f^{(1)}$
we are required to set the coefficients of Eq.(\ref{eqvinc2poli}) to zero. In other words
we have the additional restrictions,
\EQ\label{eqvinc2a}
\LAP \pm \Psi - \Psi\TPM(\dpp+\OM) = 0,
\EN
and
\EQ\label{eqvinc2b}
\pm \OM\Psi \mp \TPM(\LAP\mp \Psi) = 0.
\EN
It turns out that by solving Eq.(\ref{eqvinc2a}) for
the invariants $\TPM$ and by substituting the result 
in the companion relation (\ref{eqvinc2b}) one finds that it
is trivially satisfied once we consider the previous
constraint (\ref{eqvinc1k1}) for the invariant $\dmm$. This means that
Eqs.(\ref{eqvinc2a},\ref{eqvinc2b}) are both solved provided that we
choose,
\EQ
\label{expTPM}
\TPM = \frac{\LAP\pm\Psi}{\Psi(\dpp+\OM)}.
\EN

For the index $i=2$ we once again have to use the expressions for the weights 
$b_{-}^{(2)}$ and $c_{-}^{(2)}$ given by Eq.(\ref{expv2bmv2cm}). By substituting
these weights in 
Eq.(\ref{invTHETAPM}) 
we obtain two linear equations for the weight
$a_{-}^{(2)}$ associated to the charge sectors $\pm$. They are given by, 
\EQ\label{expv2am}
a_{-}^{(2)} = \frac{\Psi\left[a_{+}^{(2)}+\OM b_{+}^{(2)}\right]
\left\{\Psi\TPM\left[d^{(2)}\right]^2+b_{+}^{(2)}f^{(2)}\mp \TPM\left[f^{(2)}\right]^2\right\}}
{\LAP b_{+}^{(2)} \left[\TPM b_{+}^{(2)} \mp f^{(2)}\right]}.
\EN

The compatibility
of such relations for the sectors $\pm$ and the fact that the weights should be not null lead us
to the following relation among the weights 
$b_{+}^{(2)}$, $d^{(2)}$
and $f^{(2)}$, 
\EQ\label{compv2amTPM}
\left(\TM-\TP \right)\left\{\left[b_{+}^{(2)}\right]^2+\left[f^{(2)}\right]^2\right\}+
\Psi\left(\TP+\TM\right)\left[d^{(2)}\right]^2+2\left(1-\TP\TM\right)b_{+}^{(2)}f^{(2)}
= 0.
\EN
Taking into account the constraints
(\ref{eqvinc1k1},\ref{eqvinc1k2},\ref{expTPM}),
one is able to show that Eq.(\ref{compv2amTPM})
becomes equivalent to the following expression,
\EQ\label{compv2amTPM1}
\Psi^2 (\dpp+\OM) \left\{ \left[b_{+}^{(2)}\right]^2+\left[f^{(2)}\right]^2-\LAP \left[d^{(2)}\right]^2 -\dpp b_{+}^{(2)} f^{(2)}\right\}
=0.
\EN
The unique solution of Eq.(\ref{compv2amTPM1}) that does not drive us to null weights
is\footnote{Note that $\Psi = 0$ implies $\tilde{d}^{(i)} = 0$
while $\dpp+\OM = 0$  is equivalent to 
$b_{+}^{(2)}b_{-}^{(2)}
-\Psi \left[d^{(2)}\right]^2 \pm \left[f^{(2)}\right]^2 = 0$. The
latter identity implies that $f^{(2)} = 0$.},
\EQ \label{invLAPv2}
\frac{\left[b_{+}^{(2)}\right]^2+\left[f^{(2)}\right]^2-\dpp b_{+}^{(2)}f^{(2)}}{\left[d^{(2)}\right]^2} = \LAP.
\EN

Interesting enough, Eq.(\ref{invLAPv2}) generalizes the
left hand side of the invariant (\ref{invLAPM}) to the weights
with index $i=2$. At this point we observe this fact also works
for the invariant $\LAM$ involving the weights $b_{-}^{(2)}$, $d^{(2)}$ and $f^{(2)}$.
In fact, by using the expression for the weights $b_{-}^{(2)}$, 
$c_{-}^{(2)}$, 
$a_{-}^{(2)}$ coming  from
Eqs.(\ref{expv2bmv2cm},\ref{expv2am})  besides 
the constraints (\ref{constraintLAM},\ref{eqvinc1k1},\ref{eqvinc1k2},\ref{expTPM}) among
the invariants as well as Eq.(\ref{invLAPv2}) to eliminate the weight 
$[d^{(2)}]^2$ 
one is able to verify that,
\EQ \label{invLAMv2}
\frac{\left[b_{-}^{(2)}\right]^2+\left[f^{(2)}\right]^2-\dmm b_{-}^{(2)}f^{(2)}}{\left[d^{(2)}\right]^2} = \LAM.
\EN

We can now return to discuss the solution of the remaining
equations of group $G_{\pm}^{(3)}$. At this stage we have just to match
the weights 
$a_{\pm}^{(0)}$ and $b_{\pm}^{(0)}$ obtained from distinct pair of relations
associated to the groups $G_{\pm}^{(1)}$ and $G_{\pm}^{(3)}$.  
In other words, we have to make 
Eq.(\ref{expv0apv0am}) compatible with 
Eq.(\ref{expv0apv0amG3}) as well as 
Eq.(\ref{expv0bpv0bm}) consistent with 
Eq.(\ref{expv0bpv0bmG3}). This  require us to solve
the following functional relations,
\EQ\label{compv0bpv0bm}
d^{(1)}d^{(2)}\left[b_{\pm}^{(1)}a_{\pm}^{(2)}-a_{\pm}^{(1)}b_{\pm}^{(2)}\right]
\left[b_{\pm}^{(1)}b_{\pm}^{(2)}-f^{(1)}f^{(2)}\right]
= c_{\pm}^{(1)}c_{\pm}^{(2)}
\left\{b_{\pm}^{(1)}f^{(1)}\left[d^{(2)}\right]^2 - \left[d^{(1)}\right]^2 b_{\pm}^{(2)}f^{(2)}\right\},
\EN
\bear\label{compv0apv0am}
&&d^{(1)}d^{(2)}\left\{b_{\pm}^{(1)}a_{\pm}^{(2)}b_{\pm}^{(2)}-a_{\pm}^{(1)}\left[b_{\pm}^{(2)}\right]^2+
a_{\pm}^{(1)}\left[c_{\pm}^{(2)}\right]^2\right\}
\left[b_{\pm}^{(1)}b_{\pm}^{(2)}-f^{(1)}f^{(2)}\right]\nonumber\\
&&= c_{\pm}^{(1)}c_{\pm}^{(2)}a_{\pm}^{(2)}\left\{\left[b_{\pm}^{(1)}\right]^2\left[d^{(2)}\right]^2 -
\left[d^{(1)}\right]^2\left[f^{(2)}\right]^2\right\}.
\ear
Further progress are made by squaring both side of the
expressions (\ref{compv0bpv0bm},\ref{compv0apv0am}). This operation
makes it possible to use the invariants $\dpm$ (\ref{invDELTAPM}) to
eliminate the weights $\left[c_{\pm}^{(i)}\right]^2$ as well as
Eqs.(\ref{invLAPM},\ref{invLAPv2},\ref{invLAMv2}) to
extract $\left[d^{(i)}\right]^2$ for  both $i=1,2$. By performing
such two step procedure we are able to show 
that Eqs.(\ref{compv0bpv0bm},\ref{compv0apv0am})
become proportional to the expression,
\bear\label{compv0}
&&b_{\pm}^{(2)}\left[a_{\pm}^{(2)}+f^{(2)}\right]\left\{\left[b_{\pm}^{(1)}\right]^2 -a_{\pm}^{(1)}f^{(1)}\right\}
-b_{\pm}^{(1)}\left[a_{\pm}^{(1)}+f^{(1)}\right]\left\{\left[b_{\pm}^{(2)}\right]^2 -a_{\pm}^{(2)}f^{(2)} \right\}+\nonumber\\
&&-\dpm \left\{a_{\pm}^{(2)}f^{(2)} \left[b_{\pm}^{(1)}\right]^2 - a_{\pm}^{(1)}f^{(1)} \left[b_{\pm}^{(2)}\right]^2\right\} = 0.
\ear
By inspecting Eq.(\ref{compv0})  we conclude that such relation can indeed be separated leading us to our last 
invariant associated to the three terms relations,
\EQ\label{invGAMAPM}
\frac{a_{\pm}^{(i)}b_{\pm}^{(i)} - \Delta_{\pm}a_{\pm}^{(i)}f^{(i)} + b_{\pm}^{(i)}f^{(i)}}
{\left[b_{\pm}^{(i)}\right]^2 - a_{\pm}^{(i)}f^{(i)}} = \GAPM~~\mathrm{for}~~i=1,2.
\EN
where $\Gamma_{\pm}$ are additional variables.

The final step of our analysis consists in matching the
different charge sectors $\pm$ of the invariant (\ref{invGAMAPM}). 
This task involves the manipulation of
cumbersome expressions and the respective technical details
are presented in Appendix A. The condition of consistency
turns out to be a constraint among the invariants
$\GAPM$, $\LAP$ and $\OM$ whose expression is rather simple, namely
\EQ\label{constraintGAM}
\GAM\LAP - \Psi(\GAP + \OM) = 0.
\EN

In addition, an important byproduct of the analysis 
performed in Appendix A is that 
the only independent 
weights are $a_{+}^{(i)}$, $b_{+}^{(i)}$ and $c_{+}^{(i)}$. In fact, 
remaining amplitudes entering the three terms relations
can be written in terms of such weights. As a result, the weights
$a_{-}^{(i)}$, $b_{-}^{(i)}$, $c_{-}^{(i)}$,
$d^{(i)}$ and $f^{(i)}$ are given in terms of ratios of polynomials
whose degrees are at most two. In what follows we shall present such relations since
they are going to be useful later on. Following Appendix A, the simplified 
expressions for weights 
$a_{-}^{(i)}$, $b_{-}^{(i)}$, $c_{-}^{(i)}$ are,
\EQ\label{finalexpam}
a_{-}^{(i)} = \frac{\Psi^2(a_{+}^{(i)}+\OM b_{+}^{(i)})[(\dpp-\GAP+\OM)a_{+}^{(i)}-
(1+\GAP\OM)b_{+}^{(i)}]}{\LAP^2[(\dpp-\GAP)a_{+}^{(i)}-b_{+}^{(i)}]},
\EN
\EQ\label{finalexpbm}
b_{-}^{(i)} = \frac{\Psi[(\dpp-\GAP+\OM)a_{+}^{(i)}-(1+\GAP\OM)b_{+}^{(i)}]}
{\LAP[(\dpp-\GAP)a_{+}^{(i)}-b_{+}^{(i)}]}b_{+}^{(i)},
\EN
\EQ\label{finalexpcm}
c_{-}^{(i)} = \frac{\Psi^2[(\dpp-\GAP+\OM)a_{+}^{(i)}-(1+\GAP\OM)b_{+}^{(i)}]}{\LAP^2[(\dpp-\GAP)a_{+}^{(i)}-b_{+}^{(i)}]}c_{+}^{(i)},
\EN
while for weights 
$d^{(i)}$ and $f^{(i)}$ we have,
\EQ \label{dflinear}
d^{(i)} = \pm \sqrt{\frac{1-\dpp \GAP + \GAP^2}{\LAP}}\frac{b_{+}^{(i)}c_{+}^{(i)}}
{\left(-\dpp + \GAP \right)a_{+}^{(i)} + b_{+}^{(i)}},~~~
f^{(i)} = \frac{\left[a_{+}^{(i)} - \GAP b_{+}^{(i)}\right]b_{+}^{(i)}}
{\left(\dpp - \GAP \right)a_{+}^{(i)} - b_{+}^{(i)}}.
\EN

We would like to conclude this section with the following comments.
First it is important to stress that all the invariants
obtained so far are also 
valid for index $i=0$. 
This  result 
is verified by using the expressions of the
eliminated weights
$a_{\pm}^{(0)}$,
$b_{\pm}^{(0)}$,
$c_{\pm}^{(0)}$,
$d^{(0)}$ and $f^{(0)}$ on the form of the respective invariants.
As a consequence of that, the above expressions (\ref{finalexpam}-\ref{dflinear}) remain valid
for the weights 
$a_{-}^{(0)}$, $b_{-}^{(0)}$, $c_{-}^{(0)}$,
$d^{(0)}$ and $f^{(0)}$. 
We next note that out of ten possible invariants
values $\dpm$, $\LAPM$, $\GAPM$, $\TPM$, $\OM$ and $\Psi$
we end up with only four free parameters because of the six constraints
(\ref{constraintLAM},\ref{eqvinc1k1},\ref{eqvinc1k2},\ref{expTPM},\ref{constraintGAM}).
For sake of completeness we summarized such conclusions on Figure 
\ref{resumo3termos}.

\setlength{\unitlength}{1cm}
\begin{figure}[ht]
\begin{center}
\begin{picture}(10,10)
{\put(2.5,8){
\framebox(5,1.5){
$\frac{\left[a_{\pm}^{(i)}\right]^{2} + \left[b_{\pm}^{(i)}\right]^{2} - \left[c_{\pm}^{(i)}\right]^{2}}
{a_{\pm}^{(i)}b_{\pm}^{(i)}} 
= \Delta_{\pm}$
}}}
\put(5,8){\line(0,-1){0.5}}

{\put(1.5,6){
\framebox(7,1.5){
$\frac{\left[b_{\pm}^{(i)}\right]^2+\left[f^{(i)}\right]^2-\Delta_{\pm}b_{\pm}^{(i)}f^{(i)}}{\left[d^{(i)}\right]^2} =
\Psi\frac{b_{\mp}^{(i)}c_{\pm}^{(i)}}{b_{\pm}^{(i)}c_{\mp}^{(i)}} = \LAPM$
}}}
\put(5,6){\line(0,-1){0.5}}

{\put(2,4){
\framebox(6,1.5){
$\frac{\frac{\LAP}{\Psi}b_{-}^{(i)}-b_{+}^{(i)}}{f^{(i)}} =
\frac{a_{-}^{(i)}c_{+}^{(i)}-a_{+}^{(i)}c_{-}^{(i)}}{b_{+}^{(i)}c_{-}^{(i)}} = \OM$
}}}
\put(5,4){\line(0,-1){0.5}}

{\put(2.5,2){
\framebox(5,1.5){
$\frac{\left[b_{+}^{(i)}\pm b_{-}^{(i)}\right]f^{(i)}}
{b_{+}^{(i)}b_{-}^{(i)}-\Psi \left[d^{(i)}\right]^2 \pm \left[f^{(i)}\right]^2} =
\Theta_{\pm}$
}}}
\put(5,2){\line(0,-1){0.5}}

{\put(2.25,0){
\framebox(5.5,1.5){
$\frac{a_{\pm}^{(i)}b_{\pm}^{(i)} - \Delta_{\pm}a_{\pm}^{(i)}f^{(i)} + b_{\pm}^{(i)}f^{(i)}}
{\left[b_{\pm}^{(i)}\right]^2 - a_{\pm}^{(i)}f^{(i)}} = \GAPM$
}}}
\end{picture}
\end{center}
\caption{The invariant structures solving the three terms functional relations (\ref{eq2e120}-\ref{eq95e21})
valid for $i=0,1,2$ whose values are constrained by 
Eqs.(\ref{constraintLAM},\ref{eqvinc1k1},\ref{eqvinc1k2},\ref{expTPM},\ref{constraintGAM}).}
\label{resumo3termos}
\end{figure}

\subsection{Four terms relations}
After using the two terms solution (\ref{gaugetransf},\ref{invPSI})
the total number of four terms relations is reduced to fifty one equations.
The majority of these relations depend on the weights
$g^{(i)}$, $h^{(i)}$ and $\tilde{h}^{(i)}$ that are
still to be determined. The only exception is a factorizable functional 
equation given by,
\EQ \label{eqbranch}
d^{(0)}d^{(2)}\left(\Psi^2 - 1\right)\left[c_{+}^{(1)} - c_{-}^{(1)}\right] = 0.
\EN
The other fifty equations can be classified in five different groups characterized by the
number and the type of
unknowns weights $g^{(i)}$, $h^{(i)}$, $\tilde{h}^{(i)}$ present in  such functional relations.
We have equations involving only one of the
weights, the conjugate pair of amplitudes 
$h^{(i)}$ and $\tilde{h}^{(i)}$ as well as all the weights
$g^{(i)}$, $h^{(i)}$, $\tilde{h}^{(i)}$ together. In what follows we list
these distinct groups of relations,
\begin{itemize}
\item group $G_{\pm}^{(4)}$
\EQ \label{eq14e83}
-c_{\pm}^{(0)}c_{\pm}^{(1)}b_{\pm}^{(2)} + \Psi d^{(0)}d^{(1)}b_{\pm}^{(2)} +
g^{(0)}b_{\pm}^{(1)}c_{\pm}^{(2)} - b_{\pm}^{(0)}g^{(1)}c_{\pm}^{(2)} = 0,
\EN
\EQ \label{eq44e71}
- b_{\pm}^{(0)}c_{\pm}^{(1)}c_{\pm}^{(2)} + \Psi b_{\pm}^{(0)}d^{(1)}d^{(2)}
- c_{\pm}^{(0)}g^{(1)}b_{\pm}^{(2)} + c_{\pm}^{(0)}b_{\pm}^{(1)}g^{(2)} = 0,
\EN
\EQ \label{eq20e62}
- d^{(0)}b_{\mp}^{(1)}c_{\pm}^{(2)} + c_{\mp}^{(0)}b_{\pm}^{(1)}d^{(2)} -g^{(0)}d^{(1)}b_{\pm}^{(2)}
+ b_{\mp}^{(0)}d^{(1)}g^{(2)} = 0,
\EN
\item group $G_{\pm}^{(5)}$
\EQ \label{eq80}
h^{(0)}(b_{-}^{(1)}b_{+}^{(2)} - b_{+}^{(1)}b_{-}^{(2)}) + d^{(0)}d^{(1)}(c_{+}^{(2)} - c_{-}^{(2)}) = 0,
\EN
\EQ \label{eq113}
h^{(1)}(b_{+}^{(0)}b_{+}^{(2)} - b_{-}^{(0)}b_{-}^{(2)}) - c_{-}^{(0)}c_{+}^{(1)}c_{-}^{(2)} + c_{+}^{(0)}c_{-}^{(1)}c_{+}^{(2)} = 0,
\EN
\EQ \label{eq99}
h^{(2)}(b_{+}^{(0)}b_{-}^{(1)} - b_{-}^{(0)}b_{+}^{(1)}) + d^{(1)}d^{(2)}(c_{+}^{(0)} - c_{-}^{(0)}) = 0,
\EN
\EQ \label{eq112e81}
-d^{(0)}d^{(1)}b_{\mp}^{(2)} + c_{\pm}^{(0)}c_{\mp}^{(1)}b_{\pm}^{(2)} -
h^{(0)}b_{\pm}^{(1)}c_{\mp}^{(2)} + b_{\pm}^{(0)}h^{(1)}c_{\pm}^{(2)} = 0,
\EN
\EQ \label{eq101e111}
c_{\pm}^{(0)}h^{(1)}b_{\pm}^{(2)} + b_{\pm}^{(0)}c_{\mp}^{(1)}c_{\pm}^{(2)} -
b_{\mp}^{(0)}d^{(1)}d^{(2)} - c_{\mp}^{(0)}b_{\pm}^{(1)}h^{(2)} = 0,
\EN
\EQ \label{eq94e127}
a_{\pm}^{(0)}h^{(1)}a_{\pm}^{(2)} - d^{(0)}c_{\pm}^{(1)}d^{(2)} - f^{(0)}h^{(1)}f^{(2)} - h^{(0)}a_{\pm}^{(1)}h^{(2)} = 0,
\EN
\item group $G_{\pm}^{(6)}$
\EQ \label{eq50}
\tilde{h}^{(0)}(b_{+}^{(1)}b_{-}^{(2)} - b_{-}^{(1)}b_{+}^{(2)}) + \Psi^2 d^{(0)}d^{(1)}(c_{-}^{(2)} - c_{+}^{(2)}) = 0,
\EN
\EQ \label{eq17}
\tilde{h}^{(1)}(b_{-}^{(0)}b_{-}^{(2)} - b_{+}^{(0)}b_{+}^{(2)}) + c_{-}^{(0)}c_{+}^{(1)}c_{-}^{(2)}
- c_{+}^{(0)}c_{-}^{(1)}c_{+}^{(2)} = 0,
\EN
\EQ \label{eq31}
\tilde{h}^{(2)}(- b_{+}^{(0)}b_{-}^{(1)} + b_{-}^{(0)}b_{+}^{(1)}) + \Psi^2 d^{(1)}d^{(2)}(c_{-}^{(0)} - c_{+}^{(0)}) = 0,
\EN
\EQ \label{eq49e18} 
\Psi^2 d^{(0)}d^{(1)}b_{\mp}^{(2)} - c_{\pm}^{(0)}c_{\mp}^{(1)}b_{\pm}^{(2)} -
b_{\pm}^{(0)}\tilde{h}^{(1)}c_{\pm}^{(2)} + \tilde{h}^{(0)}b_{\pm}^{(1)}c_{\mp}^{(2)} = 0,
\EN
\EQ \label{eq19e29}
-c_{\pm}^{(0)}\tilde{h}^{(1)}b_{\pm}^{(2)} - b_{\pm}^{(0)}c_{\mp}^{(1)}c_{\pm}^{(2)} +
\Psi^2 b_{\mp}^{(0)}d^{(1)}d^{(2)} + c_{\mp}^{(0)}b_{\pm}^{(1)}\tilde{h}^{(2)} = 0,
\EN
\EQ \label{eq3e36}
-a_{\pm}^{(0)}\tilde{h}^{(1)}a_{\pm}^{(2)} + f^{(0)}\tilde{h}^{(1)}f^{(2)} + \Psi^2 d^{(0)}c_{\pm}^{(1)}d^{(2)}
+ \tilde{h}^{(0)}a_{\pm}^{(1)}\tilde{h}^{(2)} = 0,
\EN
\item group $G_{\pm}^{(7)}$
\EQ \label{eq25e105}
(\Psi^2-1)d^{(0)}c_{\pm}^{(1)}d^{(2)}+ \tilde{h}^{(0)}h^{(1)}\tilde{h}^{(2)} - h^{(0)}\tilde{h}^{(1)}h^{(2)} = 0,
\EN
\EQ \label{eq26e85}
\Psi h^{(0)}d^{(1)}b_{\pm}^{(2)} + d^{(0)}b_{\pm}^{(1)}c_{\pm}^{(2)} - \Psi c_{\pm}^{(0)}b_{\pm}^{(1)}d^{(2)}
- b_{\pm}^{(0)}d^{(1)}\tilde{h}^{(2)} = 0,
\EN
\EQ \label{eq45e104}
-\tilde{h}^{(0)}d^{(1)}b_{\pm}^{(2)} - \Psi d^{(0)}b_{\pm}^{(1)}c_{\pm}^{(2)} +
c_{\pm}^{(0)}b_{\pm}^{(1)}d^{(2)} + \Psi b_{\pm}^{(0)}d^{(1)}h^{(2)} = 0,
\EN
\EQ \label{eq28e125}
h^{(0)}f^{(1)}a_{\pm}^{(2)} - h^{(0)}a_{\pm}^{(1)}f^{(2)} -
\Psi d^{(0)}c_{\pm}^{(1)}d^{(2)} - f^{(0)}h^{(1)}\tilde{h}^{(2)} = 0,
\EN
\EQ \label{eq5e102}
-\tilde{h}^{(0)}f^{(1)}a_{\pm}^{(2)} + \tilde{h}^{(0)}a_{\pm}^{(1)}f^{(2)}+
\Psi d^{(0)}c_{\pm}^{(1)}d^{(2)} + f^{(0)}\tilde{h}^{(1)}h^{(2)} = 0,
\EN
\EQ \label{eq23e39}
h^{(0)}\tilde{h}^{(1)}f^{(2)} + \Psi d^{(0)}c_{\pm}^{(1)}d^{(2)} +
f^{(0)}a_{\pm}^{(1)}\tilde{h}^{(2)} - a_{\pm}^{(0)}f^{(1)}\tilde{h}^{(2)} = 0,
\EN
\EQ \label{eq91e107}
- \tilde{h}^{(0)}h^{(1)}f^{(2)} -\Psi d^{(0)}c_{\pm}^{(1)}d^{(2)}-
f^{(0)}a_{\pm}^{(1)}h^{(2)} + a_{\pm}^{(0)}f^{(1)}h^{(2)} = 0,
\EN
\item group $G_{\pm}^{(8)}$
\EQ \label{eq4e69}
-\Psi c_{\pm}^{(0)}d^{(1)}a_{\pm}^{(2)} + f^{(0)}\tilde{h}^{(1)}d^{(2)} +
\Psi d^{(0)}c_{\pm}^{(1)}g^{(2)} + \Psi \tilde{h}^{(0)}a_{\pm}^{(1)}d^{(2)} = 0,
\EN 
\EQ \label{eq61e126}
c_{\pm}^{(0)}d^{(1)}a_{\pm}^{(2)} - h^{(0)}a_{\pm}^{(1)}d^{(2)} -
d^{(0)}c_{\pm}^{(1)}g^{(2)} - \Psi f^{(0)}h^{(1)}d^{(2)} = 0,
\EN
\EQ \label{eq93e119}
a_{\pm}^{(0)}d^{(1)}c_{\pm}^{(2)} - g^{(0)}c_{\pm}^{(1)}d^{(2)} -
\Psi d^{(0)}h^{(1)}f^{(2)} - d^{(0)}a_{\pm}^{(1)}h^{(2)} = 0,
\EN
\EQ \label{eq11e37}
d^{(0)}\tilde{h}^{(1)}f^{(2)} - \Psi a_{\pm}^{(0)}d^{(1)}c_{\pm}^{(2)} +
\Psi g^{(0)}c_{\pm}^{(1)}d^{(2)} + \Psi d^{(0)}a_{\pm}^{(1)}\tilde{h}^{(2)} = 0,
\EN
\EQ \label{eq77e34}
h^{(0)}c_{\mp}^{(1)}f^{(2)} - c_{\pm}^{(0)}f^{(1)}c_{\mp}^{(2)} +
\Psi d^{(0)}g^{(1)}d^{(2)} + f^{(0)}c_{\pm}^{(1)}\tilde{h}^{(2)} = 0,
\EN
\EQ \label{eq96e53}
c_{\pm}^{(0)}f^{(1)}c_{\mp}^{(2)} - \Psi d^{(0)}g^{(1)}d^{(2)} -
\tilde{h}^{(0)}c_{\mp}^{(1)}f^{(2)} - f^{(0)}c_{\pm}^{(1)}h^{(2)} = 0.
\EN
\end{itemize}

From Eq.(\ref{eqbranch}) we see that we have to deal with at least two possible
branches. Either $\Psi^2=1$ or $c_{+}^{(1)}=c_{-}^{(1)}$ since 
the weights $d^{(i)}$ are assumed non-null.  This information, however, 
is not necessary to obtain the invariant values associated to the weights
$g^{(i)}$. Indeed, 
we observe that Eqs.(\ref{eq14e83},\ref{eq44e71}) from group $G_{\pm}^{(4)}$ 
contain the triple products
$b_{\pm}^{(0)}c_{\pm}^{(1)}c_{\pm}^{(1)}$,
$c_{\pm}^{(0)}c_{\pm}^{(1)}b_{\pm}^{(2)}$,
$\Psi d^{(0)}d^{(1)}b_{\pm}^{(2)}$ and
$\Psi b_{\pm}^{(0)}d^{(1)}d^{(2)}$
which are also present in the previously solved
three terms relations, see Eqs.(\ref{eq2e120},\ref{eq8e89},\ref{eq22e75},\ref{eq95e21}).
By eliminating the first triple products of the latter mentioned  relations
and substituting them in Eqs.(\ref{eq14e83},\ref{eq44e71}) we obtain the following
expressions,
\EQ\label{invgnotseparable1}
b_{\pm}^{(0)}\left[a_{\pm}^{(1)}-g^{(1)}\right] - \left[a_{\pm}^{(0)}- g^{(0)}\right]b_{\pm}^{(1)}
- f^{(0)}b_{\mp}^{(1)} + b_{\mp}^{(0)}f^{(1)} = 0,
\EN
and
\EQ\label{invgnotseparable2}
\left[a_{\pm}^{(1)}-g^{(1)}\right]b_{\pm}^{(2)} - b_{\pm}^{(1)}\left[a_{\pm}^{(2)}- g^{(2)}\right]
+ f^{(1)}b_{\mp}^{(2)} - b_{\mp}^{(1)}f^{(2)} = 0.
\EN
An essential step to separate Eqs.(\ref{invgnotseparable1},\ref{invgnotseparable2}) is to use
the invariant $\OM$ which links the weights $b_{+}^{(i)}$ and $b_{-}^{(i)}$. Indeed, from  the third box of
Figure \ref{resumo3termos} one is able to eliminate 
$b_{\pm}^{(i)}$ in terms of 
$b_{\mp}^{(i)}$ and $f^{(i)}$. By substituting the result in 
Eqs.(\ref{invgnotseparable1},\ref{invgnotseparable2})
and after few algebraic manipulations which includes the use of identity
(\ref{constraintLAM}) we find that the invariant associated to the weight $g^{(i)}$ is,
\EQ \label{invDELTAGPM}
\frac{-g^{(i)} + a_{\pm}^{(i)} + \frac{\Psi}{\LAPM}f}{b_{\pm}^{(i)}} = \DGPM~~\mathrm{for}~~i=0,1,2,
\EN
where $\DGPM$ are constant parameters. 

Our next task is to assure the consistency of Eq.(\ref{invDELTAGPM}) as far as the charge sectors $\pm$
are concerned. This
requires us to impose that the expressions for $g^{(i)}$ coming from
the different charge sectors are the same, namely
\EQ\label{compg}
a_{+}^{(i)} - a_{-}^{(i)} + \left(\frac{\Psi}{\LAP}-\frac{\Psi}{\LAM}\right)f^{(i)} - \DGP b_{+}^{(i)} + \DGM b_{-}^{(i)} = 0
~~\mathrm{for}~~i=0,1,2.
\EN
In order to solve Eq.(\ref{compg}) we substitute the expressions for
the weights $a_{-}^{(i)}$, $b_{-}^{(i)}$ and $f^{(i)}$ derived in previous section,
see Eqs.(\ref{finalexpam},\ref{finalexpbm},\ref{dflinear}). After
some cumbersome simplifications we find that Eq.(\ref{compg})
becomes a polynomial relation on the amplitudes $a_{+}^{(i)}$ and $b_{+}^{(i)}$ having the following 
form\footnote{
For $i=0$ the
compatibility equation (\ref{compg}) is automatically
satisfied once $A_1=A_2=A_3=0$ and provided that we take into account
the relations (\ref{constraintLAM},\ref{eqvinc1k1},\ref{eqvinc1k2}) among the invariants.},
\EQ\label{constraintg}
A_1 \left[a_{+}^{(i)}\right]^2 + A_2 a_{+}^{(i)}b_{+}^{(i)} +
A_3 \left[b_{+}^{(i)}\right]^2 = 0~~\mathrm{for}~~i=1,2.
\EN
The coefficients $A_1$, $A_2$ and $A_3$ depend solely on some combinations
of certain invariants and should vanish to assure the validity of Eq.(\ref{compg}).
Their expressions after using the constraint 
(\ref{constraintLAM})
are,
\bear
A_1& =& (\dpp - \GAP)\LAP^2 - (\dpp-\GAP+\OM)\Psi^2 \equiv 0,
\nonumber \\
A_2 &=& -\LAP^3 - [1+\DGP(\dpp-\GAP)]\LAP^2\Psi + [1+\DGM(\dpp-\GAP+\OM)]\LAP\Psi^2
\nonumber\\
&+& [1-\OM(\dpp-2\GAP+\OM)]\Psi^3 \equiv 0,
\nonumber \\
A_3& =& \DGP\Psi\LAP^2 + \GAP\LAP^3 - [\DGM+\GAP+\DGM\GAP\OM]\LAP\Psi^2 + (1+\GAP\OM)\OM\Psi^3 \equiv  0.
\label{eqA1}
\ear
The next step consists in fixing the form of invariants associated to
the Boltzmann weights $h^{(i)}$ and $\tilde{h}^{(i)}$. To make progress in
this direction we have to use explicitly the two branches data encoded in Eq.(\ref{eqbranch}).

\subsubsection{Branch 1}

This branch is chosen by setting $\Psi = \pm 1$. This constraint
allows us to eliminate the weight $\tilde{h}^{(0)}$ from Eq.(\ref{eq25e105}),
\EQ\label{expv0Zh}
\tilde{h}^{(0)} = \frac{\tilde{h}^{(1)}h^{(2)}}{h^{(1)}\tilde{h}^{(2)}}h^{(0)}.
\EN
By substituting the result (\ref{expv0Zh}) in the pair of equations (\ref{eq94e127},\ref{eq3e36})
and (\ref{eq112e81},\ref{eq49e18}) we find that their linear combination
fix a relation between $h^{(i)}$ and $\tilde{h}^{(i)}$ which is,
\EQ\label{Zhequalh}
\tilde{h}^{(i)} = h^{(i)}~~\mathrm{for}~~i=0,1,2.
\EN
As a consequence of Eq.(\ref{Zhequalh}) the equations of the group
$G_{\pm}^{(5)}$ and $G_{\pm}^{(6)}$ become the same as well as the
number of independent relations of groups $G_{\pm}^{(7)}$ and $G_{\pm}^{(8)}$
are reduced to half. At this point we can apply the same strategy used to solve the weight
$g^{(i)}$. We first eliminate the triple products of weights
$d^{(0)}b_{\pm}^{(1)}c_{\pm}^{(2)}$
and $c_{\pm}^{(0)}b_{\pm}^{(1)}d^{(2)}$ from Eqs.(\ref{eq6e82},\ref{eq43e38})
and by substituting the results in Eq.(\ref{eq26e85}) we find a separable expression
for $h^{(0)}$ and $h^{(2)}$, namely
\EQ \label{invDELTAHPMv0v2}
\frac{- \Psi h^{(0)} +  \Psi a_{\pm}^{(0)} + f^{(0)}}{b_{\pm}^{(0)}} =
\frac{- h^{(2)}+  a_{\pm}^{(2)} + \Psi f^{(2)}}{b_{\pm}^{(2)}} = \DHPM,
\EN
where $\DHPM$ are constants of separability.

Once again we have to implement the compatibility among relations
associated to different charge sectors $\pm$. This matching
is performed in the same way we did for the weight $g^{(i)}$.
As a result we find the $h^{(0)}$ and $h^{(2)}$ derived
from the charge sectors $\pm$ agree provided that the following
relations between invariants are satisfied,
\EQ\label{eqvinch1}
\DHP = \frac{1}{\LAP^2}(1+\GAP\OM)(\DHM\LAP\Psi-\OM),
\EN
\EQ\label{eqvinch2}
\OM = (\dpp-\GAP)(\LAP^2-1),
\EN
\EQ\label{eqsubbranch1}
(\LAP^2-1)\{1+(\dpp-\GAP)[(\dpp-\GAP)(\LAP^2-1)-\DHM\LAP\Psi]\} = 0.
\EN

From Eq.(\ref{eqsubbranch1}) we  observe 
that branch 1 splits in two distinct families since this relation admits 
two possible solutions. We shall denominate 
such possibilities  branches 1A and 1B. For branch 1A the value of 
$\LAP$ is fixed by,
\EQ
\LAP^2=1,
\EN
while branch 1B is defined by solving 
Eq.(\ref{eqsubbranch1})  for $\DHM$,
\EQ 
\DHM = \frac{1+(\LAP^2-1)(\dpp-\GAP)^2}{\LAP\Psi(\dpp-\GAP)}.
\EN
It turns out that for both branches we are able to manipulate Eqs.(\ref{eq112e81},\ref{eq101e111})
in order to determine the only remaining weight $h^{(1)}$. It is given
by an expression similar to that found for $h^{(0)}$
and $h^{(2)}$ which is,
\EQ \label{invDELTAHPMv1}
\frac{- h^{(1)} +  a_{\pm}^{(1)} + \Psi f^{(1)}}{b_{\pm}^{(1)}} = \DHPM.
\EN

We now have reached a point in which all the Boltzmann weights have 
been determined in terms of the invariants values and the weights 
$a_{+}^{(i)}$, $b_{+}^{(i)}$ and $c_{+}^{(i)}$. By substituting them in the four terms relations
not used so far and using the invariant (\ref{invDELTAPM}) to eliminate the weight
$[c_{+}^{(i)}]^2$ we obtain polynomial equations depending on variables  
$a_{+}^{(i)}$ 
and $b_{+}^{(i)}$ whose coefficients are functions of the invariants values. By setting
these coefficients to zero and by considering the previous 
constraints (\ref{constraintLAM},\ref{eqvinc1k1},\ref{eqvinc1k2},\ref{expTPM},\ref{constraintGAM})
as well as  Eqs.(\ref{eqA1},\ref{eqvinch1},\ref{eqvinch2},\ref{eqsubbranch1}) 
we are able to  
compute the invariants values.
It turns out that once Eqs.(\ref{eq20e62},\ref{eq4e69}) are
satisfied all the other remaining relations
involving four terms are automatically fulfilled.
The final result
of such analysis is summarized in Table \ref{tab2}. 

\begin{table}
\begin{center}
\begin{tabular}{|c|c|c|}
\hline
Invariants & Branch 1A & Branch 1B \\ \hline
$\dpp$ & free & free  \\ \hline
$\dmm$ & $\dpp$ & $\frac{-\dpp+\EPS_1\sqrt{3}\sqrt{4 - \dpp^2}}{2}$  \\ \hline
$\LAP$ & 1 & $\frac{2\EPS_1\sqrt{3}\sqrt{4 - \dpp^2}}{3\dpp+\EPS_1\sqrt{3}\sqrt{4 - \dpp^2}}$ \\ \hline
$\LAM$ & 1 & $\frac{\sqrt{3}\dpp+\EPS_1\sqrt{4 - \dpp^2}}{2\EPS_1\sqrt{4 - \dpp^2}}$ \\ \hline
$\Psi$ & 1 & 1  \\ \hline
$\OM$ & 0 & $\frac{6-3\dpp^2-\EPS_1\sqrt{3}\dpp\sqrt{4 - \dpp^2}}{3\dpp+\EPS_1\sqrt{3}\sqrt{4 - \dpp^2}}$ \\ \hline
$\GAP$ & $\dpp+\EPS_1$ & $\frac{3\dpp-\EPS_1\sqrt{3}\sqrt{4 - \dpp^2}}{6}$ \\ \hline
$\GAM$ & $\dpp+\EPS_1$ & $\frac{-3\dpp+\EPS_1\sqrt{3}\sqrt{4 - \dpp^2}}{6}$ \\ \hline
$\TP$ & $\frac{2}{\dpp}$ & $\frac{\dpp+\EPS_1\sqrt{3}\sqrt{4 - \dpp^2}}{2}$ \\ \hline
$\TM$ & 0 & $\frac{-3\dpp+\EPS_1\sqrt{3}\sqrt{4 - \dpp^2}}{6}$ \\ \hline
$\DGP$ & $\dpp-\EPS_1$ & $\frac{\EPS_1\sqrt{3}+\dpp\sqrt{4 - \dpp^2}}{\sqrt{4 - \dpp^2}}$ \\ \hline
$\DGM$ & $\dpp-\EPS_1$
& $\frac{(4 - \dpp^2)(\dpp+\EPS_1\sqrt{3}\sqrt{4 - \dpp^2})}{\EPS_1\sqrt{3}\dpp\sqrt{4 - \dpp^2}+4 - \dpp^2}$ \\ \hline
$\DHP$ & $\dpp$ & $\dpp$ \\ \hline
$\DHM$ & $\dpp$ & $\frac{\sqrt{3}\sqrt{4 - \dpp^2}-\EPS_1\dpp}{2\EPS_1}$ \\ \hline
\end{tabular}
\caption{The invariant values for branch 1  where
the discrete variable $\EPS_1 = \pm 1$.}
\label{tab2}
\end{center}
\end{table}

Here we remark that the cases $\Psi=-1$ or $\LAP=-1$ leads
us to specific invariants values which can be reproduced in the context of solution 1B and another branch that are going
to be discussed in next subsection. The technical details concerning such special situations have been
collected in Appendix B.
We finally observe that branches 1A and 1B have a unique free parameter
chosen to be $\dpp$. However, only the branch 1A is invariant
under the charge symmetry $+\leftrightarrow -$.

\subsubsection{Branch 2}

Another possibility to satisfy Eq.(\ref{eqbranch}) is to set $c_{+}^{(1)}=c_{-}^{(1)}$.
This condition together with the previous relation (\ref{finalexpcm})
lead us to a polynomial equation for the weights $a_{+}^{(1)}$
and $b_{+}^{(1)}$,
\EQ \label{branch2inv}
[(\LAP^2-\Psi^2)(\dpp-\GAP)-\OM\Psi^2]a_{+}^{(1)}+[-\LAP^2+(1+\GAP\OM)\Psi^2]b_{+}^{(1)} = 0.
\EN
The linear combination (\ref{branch2inv}) is fulfilled for arbitrary
$a_{+}^{(1)}$ and $b_{+}^{(1)}$ by imposing that its coefficients are null. This fixes
the values of the invariants $\OM$ and $\LAP$\footnote{Note
that the possible solution $\dpp = \GAP + \GAP^{-1}$
is disregarded since it leads us to $d^{(i)}=0$, see Eq.(\ref{dflinear}).},
\EQ\label{vincbranch2a}
\OM = 0~~\mathrm{and}~~\LAP = \pm \Psi.
\EN
By substituting (\ref{vincbranch2a}) back in the expressions
for the weights $a_{-}^{(i)}$, $b_{-}^{(i)}$ and $c_{-}^{(i)}$
given in Eqs.(\ref{finalexpam},\ref{finalexpbm},\ref{finalexpcm})
we find the rather simple relations,
\EQ\label{csym}
\frac{a_{-}^{(i)}}{a_{+}^{(i)}} = 1,~~\frac{b_{-}^{(i)}}{b_{+}^{(i)}} = \pm,~~\frac{c_{-}^{(i)}}{c_{+}^{(i)}} = 1
~~\mathrm{for}~~i=0,1,2.
\EN
An immediate consequence of Eq.(\ref{csym})
is that we can implement several simplifications
on the four terms equations. First, we clearly see that
Eqs.(\ref{eq80},\ref{eq113},\ref{eq99},\ref{eq50},\ref{eq17},\ref{eq31})
are automatically satisfied, besides that the number of independent
equations reduces dramatically. For groups $G_{\pm}^{(5)}$ and $G_{\pm}^{(6)}$
we just have three distinct relations while for $G_{\pm}^{(7)}$ and $G_{\pm}^{(8)}$
the number of equations is reduced to half.
We are now in position to determine the weights $h^{(i)}$ and
$\tilde{h}^{(i)}$ by using the same method that fixed $g^{(i)}$.
In fact, by eliminating the triple products 
$b_{+}^{(0)}c_{+}^{(1)}c_{+}^{(2)}$,
$c_{+}^{(0)}c_{+}^{(1)}b_{+}^{(2)}$,
$\Psi d^{(0)}d^{(1)}b_{+}^{(2)}$
and
$\Psi b_{+}^{(0)}d^{(1)}d^{(2)}$ with the help of Eqs.(\ref{eq2e120},\ref{eq8e89},\ref{eq22e75},\ref{eq95e21})
and by substituting them in Eqs.(\ref{eq112e81},\ref{eq101e111})
we are able to obtain separable functional relations for
the weight $h^{(i)}$. The solution of such relations are,
\EQ \label{invDELTAHP}
\frac{-\Psi h^{(i)} + \Psi a_{+}^{(i)} + f^{(i)}}{b_{+}^{(i)}} = \DHP~~\mathrm{for}~~i=0,1,2.
\EN

By the same token the weight $\tilde{h}^{(i)}$
can also be calculated. By using the above
explained approach but now for Eqs.(\ref{eq49e18},\ref{eq19e29})
one finds,
\EQ \label{invDELTAZHP}
\frac{-\tilde{h}^{(i)} + a_{+}^{(i)} + \Psi f^{(i)}}{b_{+}^{(i)}} = \DZHP~~\mathrm{for}~~i=0,1,2.
\EN

At this point all the Boltzmann weights for branch 2 have
been determined. The remaining task is to substitute them in
Eqs.(\ref{eq94e127},\ref{eq3e36}) as well in the relations
of groups $G_{\pm}^{(7)}$ and $G_{\pm}^{(8)}$ to
fix the invariants values. Once again we have to consider two possible branches  
since from the very beginning
Eq.(\ref{vincbranch2a}) 
has two allowed solutions.
Taking
into account all previous constraints among the invariants our final
results are summarized in Table \ref{tab3}.

\newpage

\begin{table}
\begin{center}
\begin{tabular}{|c|c|c|}
\hline
branch & Branch 2A & Branch 2B \\ \hline
$\dpp$ & free & $\EPS_1\sqrt{3}$ \\ \hline
$\dmm$ & $\dpp$ & $-\EPS_1\sqrt{3}$ \\ \hline
$\LAP$ & $\frac{2-\dpp^2+\EPS_2\dpp\sqrt{\dpp^2-4}}{2}$ & $\omega$ \\ \hline
$\LAM$ & $\frac{2-\dpp^2+\EPS_2\dpp\sqrt{\dpp^2-4}}{2}$ & $\omega$ \\ \hline
$\Psi$ & $\frac{2-\dpp^2+\EPS_2\dpp\sqrt{\dpp^2-4}}{2}$ & $-\omega$ \\ \hline
$\OM$  & 0 & 0 \\ \hline
$\GAP$ & $\EPS_1$ & $\frac{\EPS_1}{\sqrt{3}}$ \\ \hline
$\GAM$ & $\EPS_1$ & $-\frac{\EPS_1}{\sqrt{3}}$ \\ \hline
$\TP$  & $\frac{2}{\dpp}$ & 0 \\ \hline
$\TM$  & 0 & $-\frac{2\EPS_1}{\sqrt{3}}$ \\ \hline
$\DGP$ & $\dpp-\EPS_1$ & 0 \\ \hline
$\DGM$ & $\dpp-\EPS_1$ & 0 \\ \hline
$\DHP$ & 0 & 0 \\ \hline
$\DZHP$ & 0 & 0 \\ \hline
\end{tabular}
\caption{The invariant values for branch 2 where 
$\omega = \exp\left(\frac{i\pi \EPS_2}{3}\right)$ and the discrete
variables $\EPS_1=\EPS_2=\pm$. }
\label{tab3}
\end{center}
\end{table}

Note that only branch 2A has a free parameter and is invariant under charge 
conjugation.
We would like to conclude this section with the following remark.
Although the algebraic curves for the weights $h^{(i)}$
and $\tilde{h}^{(i)}$ come from different equations for
branches 1 and 2, their final expressions can be put in
a unified form. In fact, they can be written as,
\EQ \label{invDELTAHPMeZHPMgeral}
\frac{-\Psi h^{(i)} + \Psi a_{\pm}^{(i)} + f^{(i)}}{b_{\pm}^{(i)}} = \DHPM,~~
\frac{-\tilde{h}^{(i)} + a_{\pm}^{(i)} + \Psi f^{(i)}}{b_{\pm}^{(i)}} = \DZHPM~~
\mathrm{for}~~i=0,1,2,
\EN

The general 
result (\ref{invDELTAHPMeZHPMgeral}) 
recovers branch 1 by simply making the identification
$\DHPM=\DZHPM$ while
for branch 2 we have to consider Eq.(\ref{csym}) and therefore we have to impose $\frac{\DHM}{\DHP} = \frac{\DZHM}{\DZHP} =  \pm$. 
In Table \ref{tab4}  we summarized the invariants obtained
in the analysis of the four terms functional relations.
\setlength{\unitlength}{1cm}
\begin{figure}[ht]
\begin{center}
\begin{picture}(10,6)
{\put(2.5,4){
\framebox(5,1.5){
$\frac{-g^{(i)} + a_{\pm}^{(i)} + \frac{\Psi}{\LAPM}f}{b_{\pm}^{(i)}} = \DGPM$
}}}
\put(5,4){\line(0,-1){0.5}}

{\put(2.5,2){
\framebox(5,1.5){
$\frac{-\Psi h^{(i)} + \Psi a_{\pm}^{(i)} + f^{(i)}}{b_{\pm}^{(i)}} = \DHPM$
}}}
\put(5,2){\line(0,-1){0.5}}

{\put(2.5,0){
\framebox(5,1.5){
$\frac{-\tilde{h}^{(i)} + a_{\pm}^{(i)} + \Psi f^{(i)}}{b_{\pm}^{(i)}} = \DZHPM$
}}}
\end{picture}
\end{center}
\caption{The invariant structure associated with the four terms equations
valid for $i=0,1,2$. The
expressions for the invariants $\DGPM$, $\DHPM$ and $\DZHPM$
are given in Tables \ref{tab2} and \ref{tab3}.}
\label{tab4}
\end{figure}

\subsection{Five terms relations}
The number of the five terms functional equations remains unchanged after the solution
of the two terms relations. Below we present the structure of the corresponding
twenty four equations,
\EQ \label{eqg12e70}
b_{\pm}^{(0)}c_{\pm}^{(1)}b_{\pm}^{(2)} - d^{(0)}\tilde{h}^{(1)}d^{(2)} - g^{(0)}c_{\pm}^{(1)}g^{(2)}
+ c_{\pm}^{(0)}g^{(1)}c_{\pm}^{(2)} - \Psi^2d^{(0)}a_{\pm}^{(1)}d^{(2)} = 0
\EN
\EQ \label{eqg60e118}
b_{\pm}^{(0)}c_{\pm}^{(1)}b_{\pm}^{(2)} - \Psi^2 d^{(0)}h^{(1)}d^{(2)} - g^{(0)}c_{\pm}^{(1)}g^{(2)}
+ c_{\pm}^{(0)}g^{(1)}c_{\pm}^{(2)} - d^{(0)}a_{\pm}^{(1)}d^{(2)} = 0
\EN
\EQ \label{eqg24e72}
\Psi b_{\pm}^{(0)}b_{\pm}^{(1)}d^{(2)} + c_{\pm}^{(0)}d^{(1)}\tilde{h}^{(2)} - d^{(0)}c_{\pm}^{(1)}g^{(2)}
- \Psi f^{(0)}a_{\pm}^{(1)}d^{(2)}-h^{(0)}\tilde{h}^{(1)}d^{(2)} = 0
\EN
\EQ \label{eqg58e106}
b_{\pm}^{(0)}b_{\pm}^{(1)}d^{(2)} + \Psi c_{\pm}^{(0)}d^{(1)}h^{(2)} - \Psi d^{(0)}c_{\pm}^{(1)}g^{(2)}
- f^{(0)}a_{\pm}^{(1)}d^{(2)} - \Psi \tilde{h}^{(0)}h^{(1)}d^{(2)} = 0
\EN 
\EQ \label{eqg13e103}
\Psi d^{(0)}b_{\pm}^{(1)}b_{\pm}^{(2)} - \Psi d^{(0)}a_{\pm}^{(1)}f^{(2)} - g^{(0)}c_{\pm}^{(1)}d^{(2)} 
- d^{(0)}\tilde{h}^{(1)}h^{(2)} + \tilde{h}^{(0)}d^{(1)}c_{\pm}^{(2)} = 0
\EN  
\EQ \label{eqg27e117}
d^{(0)}b_{\pm}^{(1)}b_{\pm}^{(2)}  - d^{(0)}a_{\pm}^{(1)}f^{(2)}- \Psi g^{(0)}c_{\pm}^{(1)}d^{(2)}
 - \Psi d^{(0)}h^{(1)}\tilde{h}^{(2)} + \Psi h^{(0)}d^{(1)}c_{\pm}^{(2)}= 0
\EN 
\EQ \label{eqg100e79}
b_{\pm}^{(0)}c_{\mp}^{(1)}b_{\pm}^{(2)} + c_{\pm}^{(0)}h^{(1)}c_{\pm}^{(2)} - f^{(0)}c_{\mp}^{(1)}f^{(2)}
- h^{(0)}c_{\pm}^{(1)}h^{(2)} - d^{(0)}g^{(1)}d^{(2)} = 0
\EN 
\EQ \label{eqg30e51}
b_{\mp}^{(0)}c_{\pm}^{(1)}b_{\mp}^{(2)} + c_{\mp}^{(0)}\tilde{h}^{(1)}c_{\mp}^{(2)} - f^{(0)}c_{\pm}^{(1)}f^{(2)}
- \tilde{h}^{(0)}c_{\mp}^{(1)}\tilde{h}^{(2)}- \Psi^2 d^{(0)}g^{(1)}d^{(2)}  = 0
\EN 
\EQ \label{eqg63e52}
\Psi d^{(0)}b_{\pm}^{(1)}b_{\mp}^{(2)} + \Psi g^{(0)}d^{(1)}c_{\mp}^{(2)} - f^{(0)}c_{\pm}^{(1)}d^{(2)}
- \Psi d^{(0)}g^{(1)}g^{(2)} - \Psi \tilde{h}^{(0)}c_{\mp}^{(1)}d^{(2)} = 0
\EN 
\EQ \label{eqg78e67}
d^{(0)}b_{\pm}^{(1)}b_{\mp}^{(2)} + g^{(0)}d^{(1)}c_{\mp}^{(2)} - \Psi f^{(0)}c_{\pm}^{(1)}d^{(2)}
- d^{(0)}g^{(1)}g^{(2)} - h^{(0)}c_{\mp}^{(1)}d^{(2)} = 0
\EN 
\EQ \label{eqg98e66}
b_{\pm}^{(0)}b_{\mp}^{(1)}d^{(2)} - \Psi d^{(0)}c_{\mp}^{(1)}f^{(2)} +
c_{\pm}^{(0)}d^{(1)}g^{(2)} - g^{(0)}g^{(1)}d^{(2)}- d^{(0)}c_{\pm}^{(1)}h^{(2)} = 0
\EN 
\EQ \label{eqg32e64}
\Psi b_{\mp}^{(0)}b_{\pm}^{(1)}d^{(2)}-d^{(0)}c_{\pm}^{(1)}f^{(2)} + \Psi c_{\mp}^{(0)}d^{(1)}g^{(2)}
- \Psi g^{(0)}g^{(1)}d^{(2)} - \Psi d^{(0)}c_{\mp}^{(1)}\tilde{h}^{(2)} = 0
\EN 

The above relations do not involve any new Boltzmann weights. As a consequence of that, we
can substitute the weights expressions obtained in
the two previous sections in Eqs.(\ref{eqg12e70}-\ref{eqg32e64})
and search for further restrictions on the
invariants values for branches 1 and 2. Remarkably enough, 
after the three and four terms relations are solved
all the above five terms equations are automatically satisfied.
Therefore, the four families of vertex models defined by the invariants
given in Tables \ref{tab2} and \ref{tab3} are exactly integrable.

\section{Parameterization and Hamiltonian}
\label{secPARA}

We shall present here the parameterization of the weights
associated to the four distinct integrable manifolds
of previous section. As discussed in section \ref{secYB} we have
only one fundamental algebraic curve involving
the weights $a_{+}^{(i)}$, $b_{+}^{(i)}$ and $c_{+}^{(i)}$.
The corresponding curve (\ref{invDELTAPM}) can be rewritten in the form
of a conic,
\EQ\label{curva1}
\left(x_i-\frac{\dpp}{2}y_i\right)^2 - \left(\frac{\dpp^2}{4}-1\right)y_i^2 = 1,
\EN
where the new variables $x_i$ and $y_i$ are related to the weights by,
\EQ\label{defXeY}
x_i = \frac{a_{+}^{(i)}}{c_{+}^{(i)}},~~~~y_i = \frac{b_{+}^{(i)}}{c_{+}^{(i)}}.
\EN
The rational parameterization of conics involves only one spectral parameter $\la_i$
and can be done through hyperbolic functions, namely
\EQ\label{eqparamet}
x_i-\frac{\dpp}{2}y_i = \cosh(\la_i),~~~~
y_i = \frac{\sinh(\la_i)}{\sqrt{\frac{\dpp^2}{4}-1}}
\EN
For later convenience we introduce the following
definition for the invariant $\dpp$,
\EQ\label{eqdpp}
\dpp = 2\cosh(\ga).
\EN
By substituting Eq.(\ref{defXeY}) in Eq.(\ref{eqparamet}) and
after using the definition (\ref{eqdpp}) one finds,
\EQ
a_{+}^{(i)} = 1,~~~~b_{+}^{(i)} = \frac{\sinh(\la_i)}{\sinh{(\la_i+\ga)}},~~~~
c_{+}^{(i)} = \frac{\sinh(\ga)}{\sinh(\la_i+\ga)},
\EN
where $a_{+}^{(i)}$ has been fixed by freedom of an overall normalization.

Before proceeding we stress that under the above parameterization the Yang-Baxter
equation is additive as far as the spectral parameters $\la_i$ are concerned,
\EQ
\la_0=\la_1-\la_2.
\EN

The remaining weights can be determinate in terms
of $a_{+}^{(i)}$, $b_{+}^{(i)}$ and $c_{+}^{(i)}$ with the help
of Eqs.
(\ref{finalexpam},\ref{finalexpbm},\ref{finalexpcm},\ref{dflinear},\ref{invDELTAGPM},\ref{invDELTAHPMeZHPMgeral}).
In what follows we list the simplified expressions of the weights
associated to the four different integrable vertex models defined by the  invariants values given
in Tables \ref{tab2} and \ref{tab3}.
\begin{itemize}
\item Branch 1A
\end{itemize}
This vertex model falls in the family of
the factorized spin-1 $S-$matrix
found by Zamolodchikov and Fateev \cite{FZ} with the additional presence of the
discrete variables $\EPS_1=\pm 1$. We remark that the transfer matrix spectrum for $\EPS_1=+1$ can be
related to that with $\EPS_1=-1$ only for $L$ even. The respective weights are given by,
\EQ\label{pesoABC1A}
a_{\pm}^{(i)} = 1,~~~b_{\pm}^{(i)} = \frac{\sinh(\la_i)}{\sinh{(\la_i+\ga)}},~~~
c_{\pm}^{(i)} = \frac{\sinh(\ga)}{\sinh(\la_i+\ga)},
\EN
\EQ\label{pesoDZD1A}
\tilde{d}^{(i)} = d^{(i)} = \pm \frac{\sinh(\ga)\sinh(\la_i)}{\sinh(\la_i+\frac{\ga}{2}+\Bar{\ga})\sinh(\la_i+\ga)},~~~
f^{(i)} = \frac{\sinh(\la_i-\frac{\ga}{2}+\Bar{\ga})\sinh(\la_i)}{\sinh(\la_i+\frac{\ga}{2}+\Bar{\ga})\sinh(\la_i+\ga)},
\EN
\EQ\label{pesoG1A}
g^{(i)} = \frac{-2\cosh(\frac{\ga}{2}-\Bar{\ga})+\cosh(\frac{3\ga}{2}+\Bar{\ga})+\cosh(2\la_i+\frac{\ga}{2}-\Bar{\ga})}
{2\sinh(\la_i+\frac{\ga}{2}+\Bar{\ga})\sinh(\la_i+\ga)},
\EN
\EQ\label{pesoHZH1A}
\tilde{h}^{(i)} = h^{(i)} = \frac{2\cosh(\frac{\ga}{2}-\Bar{\ga})\sinh^2(\frac{\ga}{2}+\Bar{\ga})}
{\sinh(\la_i+\frac{\ga}{2}+\Bar{\ga})\sinh(\la_i+\ga)},
\EN
where
$ \Bar{\ga} = \frac{i\pi}{4}(1-\EPS_1)$. 
\begin{itemize}
\item Branch 1B
\end{itemize}
The weights of this branch were obtained
on the context of the representation of the $U_q[SU(2)]$ quantum
algebra when $q$ is at roots of unity \cite{AKU,COT,SI}. Note that this model
is not charge invariant and the expressions for the weights are,
\EQ\label{pesoABC1B}
a_{+}^{(i)} = 1,~~~b_{+}^{(i)} = \frac{\sinh(\la_i)}{\sinh{(\la_i+\ga)}},~~~
c_{+}^{(i)} = \frac{\sinh(\ga)}{\sinh(\la_i+\ga)},
\EN
\EQ\label{pesoAM1B}
a_{-}^{(i)} = \frac{\sinh(\la_i-\ga+\ga_0)\sinh(\la_i-\ga)}{\sinh(\la_i+\ga-\ga_0)\sinh(\la_i+\ga)},~~~
b_{-}^{(i)} = \frac{\sinh(\ga-\la_i)\sinh(\la_i)}{\sinh(\la_i+\ga-\ga_0)\sinh(\la_i+\ga)},
\EN
\EQ\label{pesoCM1B}
c_{-}^{(i)} = \frac{\sinh(\ga_0-\ga)\sinh(\la_i-\ga)}{\sinh(\la_i+\ga-\ga_0)\sinh(\la_i+\ga)},~~~
\tilde{d}^{(i)} = d^{(i)} = \pm \frac{\sqrt{\sinh(\ga)\sinh(\ga-\ga_0)}\sinh(\la_i)}{\sinh(\la_i+\ga-\ga_0)\sinh(\la_i+\ga)},
\EN
\EQ\label{pesoF1B}
f^{(i)} = \frac{\sinh(\la_i-\ga_0)\sinh(\la_i)}{\sinh(\la_i+\ga-\ga_0)\sinh(\la_i+\ga)},~~~
g^{(i)} = \frac{-1+\cosh(2\la_i+\ga_0)+\cosh(2\ga-\ga_0)}{2\sinh(\la_i+\ga-\ga_0)\sinh(\la_i+\ga)},
\EN
\EQ\label{pesoHZH1B}
\tilde{h}^{(i)} = h^{(i)} =
\frac{\sinh(\ga)\sinh(\ga-\ga_0)}{\sinh(\la_i+\ga-\ga_0)\sinh(\la_i+\ga)},
\EN
where
$\ga_0 = \frac{i \pi}{3}\EPS_1$.

\begin{itemize}
\item Branch 2A
\end{itemize}
Apart from the presence of the extra discrete variables $\EPS_1=\EPS_2=\pm 1$ this vertex model is equivalent to the 
so-called Izergin-Korepin $R-$matrix \cite{IK}. The transfer matrix spectrum of this model is however independent
of the parameter $\EPS_2$. The explicit form of the weights are,
\EQ\label{pesoABC2A}
a_{\pm}^{(i)} = 1,~~~b_{\pm}^{(i)} = \frac{\sinh(\la_i)}{\sinh{(\la_i+\ga)}},~~~
c_{\pm}^{(i)} = \frac{\sinh(\ga)}{\sinh(\la_i+\ga)},
\EN
\EQ\label{pesoDZD2A}
d^{(i)} = 
\mp\frac{\exp\left(\EPS_2\ga\right)\sinh(\ga)\sinh(\la_i)}{\cosh(\la_i+\frac{3\ga}{2}+\Bar{\ga})\sinh(\la_i+\ga)},~~~
\tilde{d}^{(i)} = -\exp\left(-2\EPS_2\ga\right){d}^{(i)},
\EN
\EQ\label{pesoF2A}
f^{(i)} = \frac{\cosh(\la_i+\frac{\ga}{2}+\Bar{\ga})\sinh(\la_i)}{\cosh(\la_i+\frac{3\ga}{2}+\Bar{\ga})\sinh(\la_i+\ga)},
\EN
\EQ\label{pesoG2A}
g^{(i)} = \frac{-\sinh(\frac{\ga}{2}+\Bar{\ga})-\sinh(\frac{3\ga}{2}-\Bar{\ga})+\sinh(\frac{5\ga}{2}+
\Bar{\ga})+\sinh(2\la_i+\frac{3\ga}{2}-\Bar{\ga})}
{2\cosh(\la_i+\frac{3\ga}{2}+\Bar{\ga})\sinh(\la_i+\ga)},
\EN
\EQ\label{pesoH2A}
h^{(i)} = \frac{\cosh(\la_i+\frac{3\ga}{2}+\Bar{\ga})\sinh(\la_i+\ga)-
\exp\left(2\EPS_2\ga\right)\cosh(\la_i+\frac{\ga}{2}+\Bar{\ga})\sinh(\la_i)}
{\cosh(\la_i+\frac{3\ga}{2}+\Bar{\ga})\sinh(\la_i+\ga)},
\EN
\EQ\label{pesoZH2A}
\tilde{h}^{(i)} = \frac{\cosh(\la_i+\frac{3\ga}{2}+\Bar{\ga})\sinh(\la_i+\ga)-
\exp\left(-2\EPS_2\ga\right)\cosh(\la_i+\frac{\ga}{2}+\Bar{\ga})\sinh(\la_i)}
{\cosh(\la_i+\frac{3\ga}{2}+\Bar{\ga})\sinh(\la_i+\ga)},
\EN
where as before
$ \Bar{\ga} = \frac{i\pi}{4}(1-\EPS_1)$. 

\begin{itemize}
\item Branch 2B
\end{itemize}
The nineteen vertex model defined  by the weights given 
below appears to be new in the literature.
This model violates charge
conjugation symmetry since $b_{-}^{(i)} = -b_{+}^{(i)}$. The respective weights are,
\EQ\label{pesoABC2B}
a_{\pm}^{(i)} = 1,~~b_{\pm}^{(i)} = \pm \frac{\sinh(\la_i)}{\sinh{(\la_i+\ga)}},~~
c_{\pm}^{(i)} = \frac{\sinh(\ga)}{\sinh(\la_i+\ga)},
\EN
\EQ\label{pesoDZD2B}
d^{(i)} = 
\pm\epsilon_1\epsilon_2\frac{\exp\left(\frac{i\pi\EPS_2}{3}\right)
\sinh(\la)}{2\cosh(\la-2\ga)\cosh(\la)},~~~
\tilde{d}^{(i)} = -\exp\left(\frac{i\pi\EPS_2}{3}\right){d}^{(i)},
\EN
\EQ\label{pesoG2B}
f^{(i)} = -\frac{\sinh(\la_i+2\ga)\sinh(\la_i)}{\cosh(\la_i-2\ga)\cosh(\la_i)},~~~
g^{(i)} = \frac{\cosh(2\la_i)}{2\cosh(\la_i-2\ga)\cosh(\la_i)},
\EN
\EQ\label{pesoH2B}
h^{(i)} = \frac{\cosh(\la_i-2\ga)\cosh(\la_i)+\exp\left(\frac{-i\pi\EPS_2}{3}\right)
\sinh(\la_i+2\ga)\sinh(\la_i)}{\cosh(\la_i-2\ga)\cosh(\la_i)},
\EN
\EQ\label{pesoZH2B}
\tilde{h}^{(i)} = \frac{\cosh(\la_i-2\ga)\cosh(\la_i)+\exp\left(\frac{i\pi\EPS_2}{3}\right)
\sinh(\la_i+2\ga)\sinh(\la_i)}{\cosh(\la_i-2\ga)\cosh(\la_i)},
\EN
where 
$\ga = \frac{i\pi}{2}(1-\EPS_1)+\frac{i\pi}{6}\EPS_1$ such that $\EPS_1=\EPS_2=\pm 1$.

For sake of completeness we shall also present the expressions
of the one-dimensional spin chains associated to these vertex models.
At this point we observe that the matrix representation of the weights (\ref{Lmatrix}) 
at $\la=0$ is exactly the operator of permutations on $C^3\otimes C^3$.
This property tells us that the respective Hamiltonians are given
as a sum of next-neighbor terms,
\EQ \label{hamilt}
H = J_0\sum_{i=1}^{L} H_{i, i+1},
\EN
where $J_0$ is an arbitrary normalization. 

The two-body term $H_{i, i+1}$ is obtained by the action of the
permutator on the derivative of the respective 
$\mathcal{L}^{(i)}-$operators at $\lambda_i=0$.
Its general expression in terms
of spin-1 matrices is,
\bear\label{hamiltPauli}
H_{i,i+1} &=&
J_{1}S_{i}^{+}S_{i+1}^{-} + \tilde{J}_{1}S_{i}^{-}S_{i+1}^{+} +
J_{2}\left(S_i^{+}S_i^{z}\right)\left(S_{i+1}^{-}S_{i+1}^{z}\right)
+ \tilde{J}_{2}\left(S_i^{z}S_i^{-}\right)\left(S_{i+1}^{z}S_{i+1}^{+}\right)\nonumber\\
&+&
J_{3}\left(S_i^{+}\right)^2\left(S_{i+1}^{-}\right)^2 +
\tilde{J}_{3}\left(S_i^{-}\right)^2\left(S_{i+1}^{+}\right)^2 +
J_{4}\left(S_i^{+}S_i^{z}S_{i+1}^{-}\right)
+ \tilde{J}_{4}\left(S_i^{z}S_i^{-}S_{i+1}^{+}\right)\nonumber\\
&+&
J_{5}\left(S_i^{+}S_{i+1}^{-}S_{i+1}^{z}\right)
+ \tilde{J}_{5}\left(S_i^{-}S_{i+1}^{z}S_{i+1}^{+}\right)+
h_1\left(S_{i}^{z}+S_{i+1}^{z}\right)+
h_2\left[(S_{i}^{z})^2+(S_{i+1}^{z})^2\right]\nonumber\\
&+& \delta_1S_{i}^{z}S_{i+1}^{z}+
\delta_2\left(S_{i}^{z}S_{i+1}^{z}\right)^2+
\delta_3 (S_{i}^{z})^2S_{i+1}^{z}
+ \delta_4 S_{i}^{z}(S_{i+1}^{z})^2 
\ear
where $S^{\pm}$ and $S^z$ are spin-1 matrices satisfying the $SU(2)$ algebra.

We note that in Eq.(\ref{hamiltPauli}) we have omitted any chemical
potential terms that are canceled under periodic boundary
conditions. In Table \ref{tab4H} we show the values of
the couplings entering Eq.(\ref{hamiltPauli}) for all the four branches. At this point
we remark that branch 2B Hamiltonian is a particular case of an integrable spin-1
chain discovered in \cite{ALC}. This equivalence occurs when the parameters
$t_3$ and $t_p$ defined in \cite{ALC} take the values $t_3=-1$ and $t_p=\pm\frac{\sqrt{3}}{2}$.
\begin{table}
\begin{center}
\begin{tabular}{|c|c|c|c|c|c|c|}
\hline
branch & 1A & 1B & 2A & 2B \\ \hline
$J_1$ &
$\pm\frac{1}{2\sinh(\frac{\ga}{2}+\Bar{\ga})}$ &
$\pm\frac{1}{2\sqrt{r}}$ &
$\pm\frac{1}{2\exp(\EPS_2\ga)\cosh(\frac{3\ga}{2}+\Bar{\ga})}$ &
$\mp\frac{\EPS_1 \EPS_2 \omega^2}{2}$ \\ \hline

$\tilde{J}_1$ &
$J_1$ &
$J_1$ &
$J_1$ &
$J_1$ \\ \hline

$J_2$ &
$\frac{1}{\sinh(\ga)} \mp\frac{1}{\sinh(\frac{\ga}{2}+\Bar{\ga})}$ &
$\frac{\sqrt{3}\sinh(\ga-\frac{\ga_0}{2})}{2r} \mp\frac{1}{\sqrt{r}}$ &
$\frac{1}{\sinh(\ga)} \pm\frac{\sinh(\EPS_2\ga)}{\cosh(\frac{3\ga}{2}+\Bar{\ga})}$ &
$\mp \frac{\EPS_1\EPS_2}{2}$  \\ \hline

$\tilde{J}_2$ & $J_2$ & $J_2$ & $J_2$ & $J_2$\\ \hline

$J_3$ & $-\frac{\EPS_1}{4\sinh(\ga)}$ &
$-\frac{i \sqrt{3}\EPS_1}{8r}$ & $\frac{\cosh(\frac{\ga}{2}+\Bar{\ga})}{4\sinh(\ga)\cosh(\frac{3\ga}{2}+\Bar{\ga})}$ &
$-\frac{i\sqrt{3}\EPS_1}{4}$ \\ \hline

$\tilde{J}_3$ & $J_3$ & $J_3$ & $J_3$ & $J_3$ \\ \hline

$J_4$ &
$-\frac{J_2}{2}$ & 
$-\frac{1}{2\sinh(\ga-\ga_0)}\pm\frac{1}{2\sqrt{r}}$ &
$-\frac{1}{2\sinh(\ga)} \pm\frac{\exp(-\EPS_2\ga)}{2\cosh(\frac{3\ga}{2}+\Bar{\ga})}$ & 
$-i \mp\frac{\EPS_1 \EPS_2 \omega^2}{2}$\\ \hline

$\tilde{J}_4$ &
$-\frac{J_2}{2}$ & 
$J_4$ &
$J_4$ & 
$J_4$ \\ \hline

$J_5$ &
$\frac{J_2}{2}$ &
$\frac{1}{2\sinh(\ga)} \mp\frac{1}{2\sqrt{r}}$ &
$-J_4$ & 
$-i \pm\frac{\EPS_1 \EPS_2 \omega^2}{2}$ \\ \hline

$\tilde{J}_5$ &
$\frac{J_2}{2}$ &
$J_5$ &
$-J_4$ & 
$J_5$ \\ \hline

$\delta_1$ &
$\frac{2\cosh(\ga)+\EPS_1}{2\sinh(\ga)}$ &
$0$ &
$\frac{\cosh(2\EPS_2\ga)}{2[-\EPS_1\sinh(\ga)+\sinh(2\ga)]}$ &
$-\frac{i\sqrt{3}\EPS_1}{4}$ \\ \hline

$\delta_2$ &
$-\delta_1$ &
$0$ &
$\frac{4\EPS_1\cosh(\ga)-\cosh(2\EPS_2\ga)}{2[-\EPS_1\sinh(\ga)+\sinh(2\ga)]}$ &
$3\delta_1$ \\ \hline

$\delta_3$ &
$0$ &
$0$ &
$-\frac{\sinh(2\EPS_2\ga)}{\cosh(2\EPS_2\ga)}\delta_1$ &
$i\EPS_2\sqrt{3}\delta_1$ \\ \hline

$\delta_4$ &
$0$ &
$0$ &
$\frac{\sinh(2\EPS_2\ga)}{\cosh(2\EPS_2\ga)}\delta_1$ &
$-i\EPS_2\sqrt{3}\delta_1$ \\ \hline

$h_1$ &
$0$ &
$\frac{\sinh(2\ga-\ga_0)}{2r}$ &
$0$ &
$0$ \\ \hline

$h_2$ &
$\frac{\cosh(\ga)}{\sinh(\ga)}$ & 
$-\frac{\sqrt{3}\cosh(\ga+\frac{\ga_0}{2})}{4r\sinh(\ga-\ga_0)}$ &
$\frac{[-3\EPS_1+2\cosh(\ga)]\cosh(\ga)}{-\EPS_1\sinh(\ga)+\sinh(2\ga)}$ &
$0$ \\ \hline

\end{tabular}
\caption{The Hamiltonians couplings for the four branches where
$\Bar{\ga} = \frac{i\pi}{4}(1-\EPS_1)$, $\ga_0 = \frac{i \pi \EPS_1}{3}$,
$\omega = \exp\left(\frac{i\pi\EPS_2}{3}\right)$ and 
$\EPS_1=\EPS_2= \pm 1$. The variable $r$ is defined as
$r= \sinh(\ga)\sinh(\ga-\ga_0)$ 
.} \label{tab4H}
\end{center}
\end{table}

We conclude this section with the following comment. Some of the physical properties
of the first three one-dimensional spin chains have already been investigated in 
the literature, see for instance \cite{BT,HL,WTB,CIE}. This knowlodge includes the nature
of their excitation spectrum over the anti-ferromagnetic ground state.  
Considering these results we conclude that the spectrum is massless when the free parameter
$\dpp$ belongs to the interval
$-2 < \dpp <2$. By way of contrast, outside this region the spectrum is expected to 
have a non zero gap whose value depends on $\dpp$. Remarkably enough, such change of 
physical behaviour is directly related to the  variation of the geometrical form of the
fundamental algebraic curve (\ref{curva1}). In fact, in the massless regime the corresponding curve
is closed and it is described by  
an ellipse while in the massive region the form of the respective curve is
that of an open hyperbola. This relationship between the geometric form of the principal
algebraic curve  and the nature of the excitations  suggests that
the anti-ferromagnetic regime of the fourth spin-1 chain should be gapless since in this case
$\dpp =\pm \sqrt{3}$. In next  
section we shall present evidences supporting such prediction.

\section{Exact solution of Branch 2B}
\label{secAPP}

We shall present here the exact diagonalization of the
transfer matrix associated to weights of the novel
vertex model defined as branch 2B. As usual the corresponding row-to-row
transfer matrix $T(\lambda)$ can be written as the trace
of an ordered product of $\mathcal{L}^{(i)}-$operators parameterized by the spectral
parameter $\lambda$. Recall that the 
expression for such  operators in the terms of 
Weyl matrices is given by Eq.(\ref{Lmatrix}) while the corresponding weights
$a_{+}^{(i)}$, $b_{+}^{(i)}$, $c_{+}^{(i)}$,
$d^{(i)}$, $\tilde{d}^{(i)}$, $f^{(i)}$, $g^{(i)}$,
$h^{(i)}$ and $\tilde{h}^{(i)}$
are  given in Eqs.(\ref{pesoABC2B}-\ref{pesoZH2B}).

The eigenspectrum of the transfer matrix $T(\la)$
can be determined by using the algebraic Bethe ansatz framework developed in \cite{CM}.
Here we shall present only the main results that are necessary to
unveil the thermodynamic limit properties of the spin-1 chain associated 
to the vertex model 2B. Its
transfer matrix commutes with
total spin operator $\sum_{i=1}^L S_i^{z}$ where $S_i^{z}$
denotes the azimuthal component of a spin-1 matrix acting
on the $i-$th site. Consequently, the Hilbert space
can be separated in $2L+1$ sectors labeled by an integer
number $n=L-r$ where $r=0,\pm 1,\ldots,\pm L$. We shall denote by 
$\Lambda_n(\la)$ the eigenvalue of the transfer matrix 
$T(\la)$ on a given sector $n$.
By applying the method \cite{CM}
we find that the expression for
$\Lambda_n(\la)$ is,
\bear
\Lambda_n(\la) &=& \prod_{j=1}^n \EPS_1 \frac{\sinh[\la_j-\la+\frac{i\pi\EPS_1}{12}]}{\sinh[\la_j-\la-\frac{i\pi\EPS_1}{12}]}
\nonumber \\
&+&\left[ 
\EPS_1 \frac{\sinh(\la)}{\sinh{(\la+\frac{i\pi\EPS_1}{6})}}
\right]^L 
\prod_{j=1}^n \EPS_1 \frac{\sinh[2(\la-\la_j)+\frac{i\pi\EPS_1}{2}]}{\sinh[2(\la-\la_j)-\frac{i\pi\EPS_1}{6}]}
\frac{\sinh[\la-\la_j-\frac{i\pi\EPS_1}{12}]}{\sinh[\la-\la_j+\frac{i\pi\EPS_1}{12}]}
\nonumber\\
&+&\left[
-\frac{\sinh(\la+\frac{i\pi\EPS_1}{3})\sinh(\la)}{\cosh(\la-\frac{i\pi\EPS_1}{3})\cosh(\la)}
\right]^L
\prod_{j=1}^n \EPS_1\frac{\sinh[\la-\la_j-\frac{i\pi}{2}+\frac{i\pi\EPS_1}{12}]}
{\sinh[\la-\la_j+\frac{5i\pi\EPS_1}{12}]}
\ear
provided that the rapidities $\{\la\}$ satisfy the following Bethe ansatz equations,
\EQ\label{BAeq}
\left(\frac{\sinh[\la_j+\frac{i\pi\EPS_1}{12}]}{\sinh[\la_j-\frac{i\pi\EPS_1}{12}]}\right)^L =
\prod_{k\neq j}^n\frac{\sinh[2(\la_j-\la_k)+\frac{i\pi\EPS_1}{3}]}{\sinh[2(\la_j-\la_k)-\frac{i\pi\EPS_1}{3}]}
~~~~\mbox{for} ~~ j=1,\ldots,n. 
\EN

We now have the basic ingredients to investigate the
ground state and the nature of the excitations of the
corresponding spin chain. 
The corresponding eigenvalue $E_{n}(L)$
is obtained by taking the logarithmic derivative of
$\Lambda_{n}(\lambda)$ at $\lambda = 0$ and using as normalization $J_0=-i$. The result is, 
\EQ\label{eigenvalues}
E_{n}(L) = \EPS_1\sum_{j=1}^{n} \frac{2 \sin(\frac{\pi}{6})}{\cos(\frac{\pi}{6})-\cosh(2\lambda_j)}
\EN

\begin{figure}[h]
\centering
\includegraphics[width=10cm,height=10cm,angle=-90]{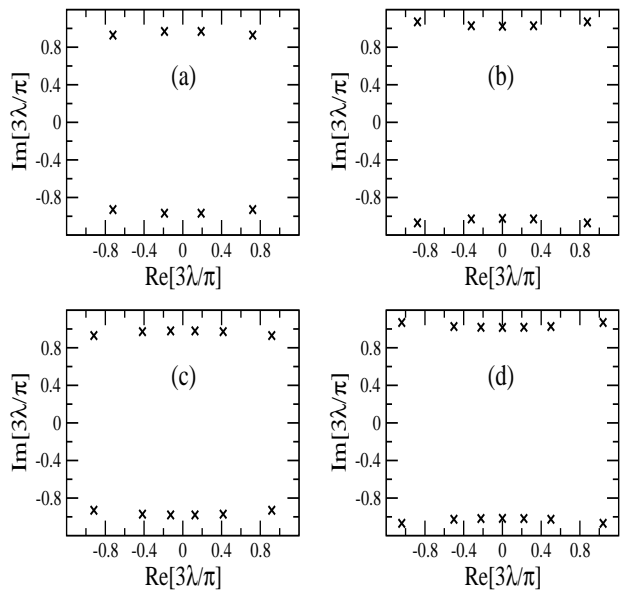}\\
\caption{The ground state roots $z_j=\frac{3\la_j}{\pi}$ for (a) $L=8$, (b) $L=10$,
(c) $L=12$, (d) $L=14$.}\label{graficodata1}
\end{figure}

From now on we shall concentrate our attention to study
the properties of the
anti-ferromagnetic regime $\EPS_1=+1$. Further progress can be made
after the identification
of the distribution of roots $\{\lambda_j\}$ that are
able to reproduce the low-lying energies of the spin
Hamiltonian. This step is done by numerically solving
the Bethe ansatz equation (\ref{BAeq}) and substituting
the roots in Eq.(\ref{eigenvalues}). We then compare the results
for $E_{n}(L)$ with the exact diagonalization of the
Hamiltonian up to $L=12$. We observe that though the Hamiltonian is not
hermitian we found that its eigenvalues are all real. 
By performing this analysis
we find that the ground state for $L$ even sits on the sector
with zero magnetization $n = L$. Interesting enough, the
shape of the roots on the complex plane depends whether
$\frac{L}{2}$ is even or odd. In Figure \ref{graficodata1}
we exhibit the
Bethe roots for $L = 8,10,12~\text{and}~14$ scaled
by the factor
$\frac{3}{\pi}$.

\begin{figure}[h]
\centering
\includegraphics[width=10cm,height=10cm,angle=-90]{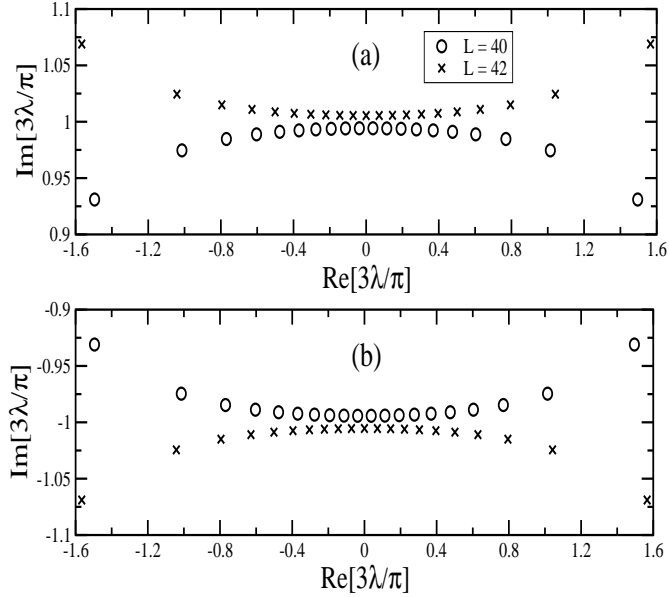}\\
\caption{The ground state roots $z_j= \frac{3\la_j}{\pi}$ for $L=40$ and $L=42$ where
(a) $z_j= x_j+iy_j$ and (b) $z_j=x_j-iy_j$.}
\label{graficodata2}
\end{figure}

We then solve the Bethe equations (\ref{BAeq}) for larger values
of $L$ to figure out the pattern of the roots in the
 thermodynamic limit $L\rightarrow\infty$. In Figure \ref{graficodata2}
we show the corresponding Bethe roots for $L=40~\mathrm{and}~42$. For better
display of the roots curvature we shown the positive and the negative imaginary
parts separately.

The above analysis leads us to conclude that as $L\rightarrow\infty$
the roots $\lambda_j$ cluster in a complex $2-$strings.
Each $2-$string has the same real part $\mu_s$ and equally
spaced imaginary parts,
\EQ\label{2strings}
\lambda_j = \mu_j \pm \frac{i\pi}{3}~,~\mu_j \in \mathbb{R}.
\EN
We now substitute the structure (\ref{2strings}) in the Bethe equations (\ref{BAeq})
and by taking its logarithm we find that the resulting
relations for $\mu_j$ are,
\EQ\label{relformu}
L\left[\phi\left(\mu_j,\frac{5\pi}{12}\right)-
\phi\left(\mu_j,\frac{\pi}{4}\right)\right]=
-2\pi Q_j + \sum_{k\neq j}^{n}\phi\left(2\mu_j-2\mu_k,\frac{\pi}{3}\right)~,
~\text{for}~j=1,\ldots,n
\EN
where the function $\phi(x,y) = 2\arctan[\tanh(x)\cot(y)]$
and $Q_j$ are integer or semi-integer numbers characterizing
the logarithm branches.

Considering our numerical analysis we find that the low-lying spectrum
can be described by the following sequence of $Q_j$ numbers,
\EQ
Q_j = -\frac{1}{2}\left[\frac{L}{2}-n-1\right]+j-1~,
~\text{for}~j=1,\ldots,\frac{L}{2}-n.
\EN
For large $L$, the roots approach toward a continuous distribution
with density $\rho(\mu)$ given by,
\EQ\label{eqrho}
\rho(\mu) = \frac{d}{d\lambda}\left[Z_L(\mu)\right]
\EN
where the counting function is,
\EQ\label{eqZL}
Z_L(\mu_j)=\frac{1}{2\pi}\left[\phi\left(\mu_j,\frac{\pi}{4}\right)-
\phi\left(\mu_j,\frac{5\pi}{12}\right)+\frac{1}{L}
\sum_{k\neq j}^{L}\phi(2\mu_j-2\mu_k,\frac{\pi}{3})\right].
\EN
Strictly in the $L\rightarrow\infty$, Eqs.(\ref{eqrho},\ref{eqZL}) go
into a linear integral equation for the density $\rho(\mu)$
which can be solved by standard Fourier transform
method. The final result for $\rho(\mu)$ is,
\EQ\label{eqfinalrho}
\rho(\mu) = \frac{1}{\pi \cosh(2\mu)}.
\EN
The ground state energy per site
$e_{\infty}$
can now be computed by taking the infinity volume limit
of Eq.(\ref{eigenvalues}) and with the help of Eq.(\ref{eqfinalrho}). By writing the result
in terms of its Fourier transform we find,
\EQ
e_{\infty} = \int_{0}^{\infty}d\omega
\frac{\sinh[\frac{\pi\omega}{12}]-\sinh[\frac{\pi\omega}{4}]}
{\cosh[\frac{\pi\omega}{4}]\sinh[\frac{\pi\omega}{2}]}=
-\frac{2}{\pi}+\frac{\sqrt{3}}{9}.
\EN

Let us turn our attention to the
behaviour of the low-lying excitations. These states are
obtained from the Bethe equations (\ref{relformu}) by introducing vacancies
on the sequence of numbers $Q_{j}$. This method is by now
standard to integrable models and for technicalities see for
instance \cite{EX1}. It turns out that the energy $\epsilon(\mu)$
and the momenta $\rho(\mu)$, measured from the ground state,
of a hole excitation are given by,
\EQ\label{eqeprho}
\epsilon(\mu) = 2\pi\rho(\mu)~~\text{and}~~\rho(\mu)=
\int_{\mu}^{\infty}\epsilon(x)dx.
\EN
In order to calculate the low-lying dispersion relation
$\epsilon(p)$ we have first to compute the integral in Eq.(\ref{eqeprho})
and afterwards eliminate the variable $\mu$ from function
$\epsilon(\mu)$. By performing these steps we find
that the dispersion relation is
\EQ\label{eqdisp}
\epsilon(p)=2\sin(p).
\EN
From Eq.(\ref{eqdisp}) we conclude that the low momenta
excitations have a linear behaviour on the respective
momenta. Therefore, the excitations are massless
corroborating the connection between the geometric form of the curve and the
nature of the spin-1 chain excitations
made at end of section \ref{secPARA}. 

\section{Conclusion}
\label{secCON}

In conclusion, we have investigated the solutions of the Yang-Baxter equation
for nineteen vertex models invariant by the action of parity-time reversal
symmetry. We have developed a method to solve the corresponding functional
equations from an algebraic point of view without the need of a priori spectral
parameterization assumption. The structure of the algebraic manifolds constraining
the Boltzmann weights follows a rather universal pattern allowing us to classify
the possible integrable vertex models in four different families. They share the same
fundamental algebraic curve which turns out to be a conic depending on three basic
weights. By using the standard parameterization of conics we are able to obtain
the dependence of the Boltzmann weights on the spectral parameter from a unified
perspective. Three of such vertex models were already known before but the fourth
model appears to be new in the literature.

We have observed an intriguing relation between the  form
of the main algebraic curve and the
nature of the low-lying excitations of the related spin-1 chains. In the regime in which
the geometric form of such curve is an ellipse the excitations are massless while
when we have a hyperbolic structure the excitations  have a mass gap. This fact is supported
by previous knowledge on the physical properties of the first three spin-1 chains and also
by the exact solution of the novel nineteen vertex model.

It seems interesting to investigate whether or not the scenario described above remains
valid for other families of $\cal{PT}$ invariant vertex models with larger number
of states. In principle, the systematic method developed here can be extended to tackle
other models  whose statistical configurations preserves at least one $U(1)$ symmetry.
Another interesting problem is to study the algebraic invariants underlying vertex models
that are not invariant by the $\cal{PT}$ symmetry and in particular to unveil the
form of their principal algebraic curve. We hope to address these questions in future
publications.

\addcontentsline{toc}{section}{Appendix A}
\section*{\bf Appendix A: Three terms relations}
\setcounter{equation}{0}
\renewcommand{\theequation}{A.\arabic{equation}}

This appendix is devoted to the presentation of
some technical details entering the solution
of the group $G_{\pm}^{(3)}$ relations.
An immediate consequence of the last invariant
(\ref{invGAMAPM}) is that the weight $f^{(i)}$
can be easily written either in terms
of the weights $a_{+}^{(i)}$ and $b_{+}^{(i)}$
or by means of the weights $a_{-}^{(i)}$ and $b_{-}^{(i)}$,
namely
\EQ \label{flineari}
f^{(i)} = \frac{\left[a_{\pm}^{(i)} - \GAPM b_{\pm}^{(i)}\right]b_{\pm}^{(i)}}
{\left(\dpm - \GAPM \right)a_{\pm}^{(i)} - b_{\pm}^{(i)}}~~\mathrm{for}~~i=1,2.
\EN

It turns out that Eq.(\ref{flineari}) can be used
to obtain a similar expression for the weight $d^{(i)}$.
In fact, by substituting Eq.(\ref{flineari}) in
Eqs.(\ref{invLAPM},\ref{invLAPv2},\ref{invLAMv2})
and by carrying on some simplification one finds,
\EQ\label{dlineardetail}
\LAPM \left[d^{(i)}\right]^2 = \frac{\left(1-\dpm \GAPM + \GAPM^2\right)
\left[b_{\pm}^{(i)}\right]^2
\left\{\left[a_{\pm}^{(i)}\right]^2-\dpm a_{\pm}^{(i)} b_{\pm}^{(i)}+\left[b_{\pm}^{(i)}\right]^2\right\}}
{\left[\left(-\dpm + \GAPM \right)a_{\pm}^{(i)} + b_{\pm}^{(i)}\right]^2}~~\mathrm{for}~~i=1,2.
\EN
By using the form of the invariants $\dpm$ given
by Eq.(\ref{invDELTAPM}) in the expression (\ref{dlineardetail})
one is able to take the square root of the weight $\left[d^{(i)}\right]^2$. The final result is,
\EQ \label{dlineari}
d^{(i)} = \pm \sqrt{\frac{1-\dpm \GAPM + \GAPM^2}{\LAPM}}\frac{b_{\pm}^{(i)}c_{\pm}^{(i)}}
{\left(-\dpm + \GAPM \right)a_{\pm}^{(i)} + b_{\pm}^{(i)}}~~\mathrm{for}~~i=1,2.
\EN

This means that the compatibility among the invariants
$\GAPM$ is equivalent to the matching of the expressions
for the weights $f^{(i)}$ and $d^{(i)}$
coming from the distinct charge sectors $\pm$.
By imposing such consistency one finds that
Eq.(\ref{flineari}) requires us to solve,
\bear\label{compfi}
\left[\left(\dmm - \GAM \right)a_{-}^{(i)} - b_{-}^{(i)}\right]
\left[a_{+}^{(i)} - \GAP b_{+}^{(i)}\right]b_{+}^{(i)}-
\left[\left(\dpp - \GAP \right)a_{+}^{(i)} - b_{+}^{(i)}\right]
\left[a_{-}^{(i)} - \GAM b_{-}^{(i)}\right]b_{-}^{(i)}
&=& 0 \nonumber \\ \mathrm{for}~~i=1,2
\ear
while Eq.(\ref{dlineari}) implies that the compatibility
for $d^{(i)}$ is,
\bear\label{compdi}
&&\sqrt{\LAM\left(1-\dpp \GAP + \GAP^2\right)}
\left[\left(-\dmm + \GAM \right)a_{-}^{(i)} + b_{-}^{(i)}\right]
b_{+}^{(i)}c_{+}^{(i)}-\nonumber\\
&&\sqrt{\LAP\left(1-\dmm \GAM + \GAM^2\right)}
\left[\left(-\dpp + \GAP \right)a_{+}^{(i)} + b_{+}^{(i)}\right]
b_{-}^{(i)}c_{-}^{(i)}
= 0~~\mathrm{for}~~i=1,2.
\ear
The analysis of Eqs.(\ref{compfi},\ref{compdi}) for the index
$i=1$ is immediate since the weights $a_{-}^{(1)}$ and $c_{-}^{(1)}$
are still free. Therefore, Eqs.(\ref{compfi},\ref{compdi})
for $i=1$ are easily solved by eliminating
the weights $a_{-}^{(1)}$ and $c_{-}^{(1)}$, namely
\EQ\label{expv1am}
a_{-}^{(1)} = \frac{\GAM(\dpp-\GAP)a_{+}^{(1)}b_{-}^{(1)}-a_{+}^{(1)}b_{+}^{(1)}-
\GAM b_{-}^{(1)}b_{+}^{(1)}+\GAP \left[b_{+}^{(1)}\right]^2}
{(\dpp-\GAP)a_{+}^{(1)}b_{-}^{(1)}-(\dmm-\GAM)a_{+}^{(1)}b_{+}^{(1)}-b_{-}^{(1)}b_{+}^{(1)}+
\GAP(\dmm-\GAM)\left[b_{+}^{(1)}\right]^2}b_{-}^{(1)}
\EN
and
\EQ\label{expv1cm}
c_{-}^{(1)} = \frac{-\sqrt{\LAM(1-\dmm\GAM+\GAM^2)(1-\dpp\GAP+\GAP^2)}b_{-}^{(1)}b_{+}^{(1)}c_{+}^{(1)}}
{\sqrt{\LAP}\left\{(\dpp-\GAP)a_{+}^{(1)}b_{-}^{(1)}-(\dmm-\GAM)a_{+}^{(1)}b_{+}^{(1)}-b_{-}^{(1)}b_{+}^{(1)}
+\GAP(\dmm-\GAM)\left[b_{+}^{(1)}\right]^2\right\}}.
\EN

On the other hand the solution of Eqs.(\ref{compfi},\ref{compdi})
for the index $i=2$ is more involving. This is the case because
the weights $a_{-}^{(2)}$, $b_{-}^{(2)}$ and $c_{-}^{(2)}$
have already been determined in terms of the amplitudes
$a_{+}^{(2)}$, $b_{+}^{(2)}$, $c_{+}^{(2)}$, $f^{(2)}$ and $d^{(2)}$
thanks to the previous relations (\ref{expv2bmv2cm},\ref{expv2am}).
In addiction to that we also recall that the weights
$f^{(2)}$ and $d^{(2)}$ are also
determined in terms of $a_{+}^{(2)}$, $b_{+}^{(2)}$ and $c_{+}^{(2)}$
through Eqs.(\ref{flineari},\ref{dlineari}).
Considering all these information together with the fact
that the weight $\left[c_{+}^{(2)}\right]^2$
can be eliminated with the help of the invariant 
$\dpp$ (\ref{invDELTAPM}) one concludes
that Eqs.(\ref{compfi},\ref{compdi})
are in fact a polynomial relation on the remaining
weights $a_{+}^{(2)}$ and $b_{+}^{(2)}$. It turns out
the expression of this polynomial
associated to Eq.(\ref{compfi}), after using the constraints (\ref{eqvinc1k1},\ref{expTPM}), is
\EQ\label{poliGAM}
B_1 \left[a_{+}^{(2)}\right]^3+
B_2 a_{+}^{(2)}\left[b_{+}^{(2)}\right]^2+
B_3 \left[a_{+}^{(2)}\right]^2 b_{+}^{(2)}+
B_4 \left[b_{+}^{(2)}\right]^3 = 0,
\EN
where its coefficients are given only
in terms of the invariants values by,
\bear\label{eqB1}
B_1 &=& \Psi[\LAP^2\GAM(- \dpp + \GAP) + 
       \LAP\Psi(\dpp\GAP + \GAM\GAP - \GAP^2+ 
       \dpp\OM+\GAM\OM - \GAP\OM) \nonumber\\&-& 
       \Psi^2 (\GAP^2 + \dpp\OM + 2\GAP\OM + 
       2\OM^2)],
\ear
\bear
\label{eqB2}
B_2 &=& \LAP^3\GAP(\dpp + 2\GAM - \GAP) + 
     \LAP^2\Psi(1 - 2\dpp\GAM - \dpp\GAP + 
     \GAM\GAP -  4\GAP\OM) + \LAP\Psi^2(-1 \nonumber\\&-& 
     \dpp\GAM + \dpp\GAP - 
     \GAM\GAP + \dpp^2\GAM\GAP +
     \dpp\OM + \GAM\OM - \GAP\OM - 
     \dpp^2\GAP\OM - \dpp\GAM\GAP\OM \nonumber\\&+& 
     \dpp\GAP^2\OM - \OM^2 - 
     \dpp\GAP\OM^2 - 2\GAM\GAP\OM^2 + 
     \GAP^2\OM^2) - \Psi^3(1 - 2\dpp\GAP + 
     \GAP^2 + \dpp^2\GAP^2 \nonumber\\&-& 2\GAP\OM +
     \dpp\GAP^2\OM + 2\OM^2 - 
     4\dpp\GAP\OM^2 - 4\GAP\OM^3),
\ear
\bear
B_3 &=& -\LAP^3(\dpp + \GAM - \GAP) + \LAP^2\Psi(\GAM + 
     \dpp^2\GAM - \GAP - \dpp\GAM\GAP + 2\OM) + \LAP\Psi^2(\dpp + 2\GAM \nonumber\\ &-& 
     2\GAP - \dpp^2\GAP - 
     2\dpp\GAM\GAP + \dpp\GAP^2 -\OM - \GAM\GAP\OM + 
     \dpp\OM^2 + \GAM\OM^2 - 
     \GAP\OM^2) \nonumber\\ &+& \Psi^3(2\dpp\GAP^2 - 3\OM +
     4\dpp\GAP\OM + \GAP^2\OM - 
     \dpp\OM^2 + 2\GAP\OM^2 - 2\OM^3),
\ear
\bear
\label{eqB4}
B_4 &=& -\LAP^3\GAP(1 + \GAM\GAP) + \LAP^2\Psi(\GAM - 
     \GAP + \dpp\GAP^2  + 2\GAP^2\OM) + 
     \LAP\Psi^2(\GAM - \dpp\GAM\GAP \nonumber\\ &+&  
     \GAM\GAP^2 - \OM + \dpp\GAP\OM
      - \GAM\GAP\OM + \dpp\GAM\GAP^2\OM
       + \GAP\OM^2 + \GAM\GAP^2\OM^2) - 
     \Psi^3(\OM - 2\dpp\GAP\OM \nonumber\\ &+& \GAP^2\OM + 
     \dpp^2\GAP^2\OM\Psi^3 - 2\GAP\OM^2\Psi^3 + 
     3\dpp\GAP^2\OM^2\Psi^3 + 2\GAP^2\OM^3\Psi^3).
\ear

As argued in the main text we are searching for solutions in which 
$a_{+}^{(2)}$ and $b_{+}^{(2)}$ are independent of each other.
This means that we have to set all the 
coefficients of the polynomial (\ref{poliGAM}) to zero. By imposing this condition for $B_1$
one is able to eliminate the invariant $\GAM$ whose expression can be further
simplified with the help of the constraint  
(\ref{eqvinc1k2}). The final result is,
\EQ\label{eqGAM}
\GAM = \frac{\Psi}{\LAP}\left(\GAP+\OM\right).
\EN

By substituting the result (\ref{eqGAM}) in the expressions of the
remaining coefficients (\ref{eqB2}-\ref{eqB4}) and by once again taking into
account the constraint 
(\ref{eqvinc1k2}) we find that  
$B_2$, $B_3$ and $B_4$ vanish. The same reasoning can be repeated for Eq.(\ref{compdi}) when 
$i=2$. We conclude that it does not impose additional restriction besides Eq.(\ref{eqGAM}).

We conclude by observing that the weights 
$a_{-}^{(i)}$, $b_{-}^{(i)}$ and $c_{-}^{(i)}$ can also be written in terms of
the amplitudes 
$a_{+}^{(i)}$, $b_{+}^{(i)}$ and $c_{+}^{(i)}$. In order to see that we first
substitute the weights 
$f^{(i)}$ e $d^{(i)}$, considering the index
$+$ of Eqs.(\ref{compfi},\ref{compdi}), in the expressions of the weights
$b_{-}^{(1)}$,
$b_{-}^{(2)}$, $c_{-}^{(2)}$, $a_{-}^{(2)}$, $a_{-}^{(1)}$ and $c_{-}^{(1)}$,
see Eqs.(\ref{qv1dpEv1bm},\ref{expv2bmv2cm},\ref{expv2am},\ref{expv1am},\ref{expv1cm}).
We next use the constraint  
$\dpp$ (\ref{invDELTAPM}) to eliminate the amplitude $\left[c_{+}^{(i)}\right]^2$ besides the explicit form of
the relations between the invariants 
(\ref{constraintLAM},\ref{eqvinc1k1},\ref{eqvinc1k2},\ref{expTPM},\ref{eqGAM}). By performing
such steps we are able to write rather simple expressions for the weights
$a_{-}^{(i)}$, $b_{-}^{(i)}$, $c_{-}^{(i)}$ for $i=1,2$. They have given in the main text,
see Eqs.(\ref{finalexpam}-\ref{finalexpcm}).

\addcontentsline{toc}{section}{Appendix B}
\section*{\bf Appendix B: Particular invariant solutions}
\setcounter{equation}{0}
\renewcommand{\theequation}{B.\arabic{equation}}
We shall describe particular solutions for the invariants coming from
branch 1A. Besides
$\Psi=\LAP=1$ we have two other possibilities that provide us non-null
Boltzmann weights. We shall denominate such special branches as follows,
\EQ
\label{eqapB1}
\bullet ~~ \mathrm{Branch~1S}:~~~\Psi=-\LAP=1,
\EN
\EQ
\label{eqapB2}
\bullet ~~ \mathrm{Branch~2S}:~~~\Psi=\LAP=-1.
\EN

It turns out that the choices 
(\ref{eqapB1},\ref{eqapB2}) lead us to the situation in which we do not
have any free parameter at our disposal. In Table \ref{tab6} we present
the corresponding invariant values.
\begin{table}
\begin{center}
\begin{tabular}{|c|c|c|}
\hline
Invariants & Branch 1S & Branch 2S  \\ \hline
$\dpp$ & $\pm \sqrt{3}$ & $\pm 2$  \\ \hline
$\dmm$ & $\mp \sqrt{3}$ & $\pm 2$  \\ \hline
$\LAP$ & $-1$ & $-1$  \\ \hline
$\LAM$ & $-1$ & $-1$  \\ \hline
$\Psi$ & $1$ & $-1$  \\ \hline
$\OM$ & $0$ & $0$  \\ \hline
$\GAP$ & $\pm\frac{2}{\sqrt{3}}$ & $\mp 1$  \\ \hline
$\GAM$ & $\mp\frac{2}{\sqrt{3}}$ & $\mp 1$  \\ \hline
$\TP$ & $0$ & $\pm 1$  \\ \hline
$\TM$ & $\mp\frac{2}{\sqrt{3}}$ & $0$  \\ \hline
$\DGP$ & $0$ & $\pm 3$  \\ \hline
$\DGM$ & $0$ & $\pm 3$  \\ \hline
$\DHP$ & $\pm\sqrt{3}$ & $0$  \\ \hline
$\DHM$ & $\mp\sqrt{3}$ & $0$  \\ \hline
$\DZHP$ & $\pm\sqrt{3}$ & $0$  \\ \hline
$\DZHM$ & $\mp\sqrt{3}$ & $0$  \\ \hline
\end{tabular}
\caption{Invariants values for the special branches 1S and 2S.
}
\end{center}
\end{table}\label{tab6}

By comparing Table \ref{tab6} with Table \ref{tab2} it is not difficult to
see that the branch 1S is a particular case of branch 1B once we choose
$\dpp = \pm \sqrt{3}$. By the same token comparison with Table \ref{tab3}
reveals us that branch 2S becomes equivalent to branch 2A  provided 
we set $\dpp = \pm 2$.

\section*{Acknowledgments}
The authors thank the Brazilian Research Agencies FAPESP and CNPq for financial support.

\end{document}